\documentclass[pageno]{jpaper}

\usepackage[normalem]{ulem}
\usepackage{authblk}

\usepackage{algorithm}
\usepackage{algpseudocode}

\usepackage{placeins}
\usepackage{multirow}
\usepackage{xspace}
\usepackage{graphicx}

\newcommand{\ptp}{Telescope\xspace}

\newcommand{\damon}{DAMON\xspace}
\newcommand{\hemem}{PMU\xspace}

\newcommand{\sota}{state-of-the-art\xspace}

\newcommand{\memcached}{Memcached\xspace}
\newcommand{\redis}{Redis\xspace}
\newcommand{\memtier}{memtier\xspace}
\newcommand{\ycsb}{YCSB\xspace}

\LetLtxMacro\origcite\cite

\newcommand{\accessed}{\texttt{ACCESSED}\xspace}
\newcommand{\masim}{MASIM\xspace}

\newcommand{\sepblock}{\noindent \leavevmode \newline}

\usepackage{pifont} 
\renewcommand{\one}{\ding{182}\xspace}
\newcommand{\two}{\ding{183}\xspace}
\newcommand{\three}{\ding{184}\xspace}

\usepackage{array}
\newcolumntype{L}[1]{>{\raggedright\let\newline\\\arraybackslash\hspace{0pt}}m{#1}}
\newcolumntype{C}[1]{>{\centering\let\newline\\\arraybackslash\hspace{0pt}}m{#1}}
\newcolumntype{R}[1]{>{\raggedleft\let\newline\\\arraybackslash\hspace{0pt}}m{#1}}

\usepackage[table]{xcolor}
\definecolor{LightCyan}{RGB}{211,211,211}

\begin{document}

\title{\ptp: Telemetry at Terabyte Scale}


\author{Alan Nair\thanks{Work done as an intern at Intel Labs. Currently at The University of Edinburgh}}
\author{Sandeep Kumar}
\author{Aravinda Prasad}
\author{Andy Rudoff\thanks{Work done while at Intel Labs}}
\author{Sreenivas Subramoney}
\affil{Processor Architecture Research Lab, Intel Labs}



\date{}
\maketitle

\thispagestyle{empty}

\begin{abstract}

Data-hungry applications that require terabytes of memory have become widespread  in recent years. 
To meet the memory needs of these applications, data centers are embracing tiered memory architectures with near and far memory tiers.  
Precise, efficient, and timely identification of hot and cold data and their placement in appropriate tiers is critical for performance in such systems.
Unfortunately, the existing state-of-the-art telemetry techniques for hot and cold data detection are ineffective at terabyte scale.

We propose \ptp, a novel technique that profiles different levels of the application's page table tree for fast and efficient identification of hot and cold data. 
\ptp is based on the observation that for a memory and TLB-intensive workload, higher levels of a page table tree are also frequently accessed during a hardware page table walk. 
Hence, the hotness of the higher levels of the page table tree essentially captures the hotness of its subtrees or address space sub-regions at a coarser granularity.
We exploit this insight to quickly converge to even a few megabytes of hot data and efficiently identify several gigabytes of cold data in terabyte-scale applications.
Importantly, such a technique can seamlessly scale to petabyte-scale applications.

\ptp's telemetry achieves 90\%$+$ precision and recall at just 0.009\% single CPU utilization for microbenchmarks with 5\,TB memory footprint. Memory tiering based on \ptp results in 5.6\% to~34\% throughput improvement for real-world benchmarks with 1--2\,TB memory footprint compared to other state-of-the-art telemetry techniques. 

\end{abstract} 

\section{Introduction}
\label{sec:intro}

The rise of big data applications has resulted in an exponential increase in data volume being generated and processed. 
The memory footprints of applications in fields such as analytics, machine learning, databases, and high-performance computing exceed petabytes in size~\cite{graph500, fb-database,scaling-ml}. For example, Meta’s database solutions mine information from geographically distributed databases
spanning petabytes in size~\cite{fb-database} and genome-sequencing workloads operate on in-memory data sets that span terabytes~\cite{genomic-sequencing}.

Increasing the DRAM memory capacity of data center servers to accommodate the needs of big data applications is not a viable solution for two reasons. First, 
memory cost has already surpassed compute cost and now accounts for up to 50\% of server cost in data centers~\cite{cxl-genz, tpp}. Increasing the DRAM capacity further can escalate the total cost of ownership (TCO) and can directly impact cloud or data center economics. Second, prior studies in cloud data centers have shown that more than half of the data is not frequently accessed and hence are cold data~\cite{alibaba-analysis-clustering, google-trace, fmr, hpc-memory-underutilization, imbalance-cloud}.
Using costly DRAM memory for storing infrequently accessed cold data is imprudent. Storing
the cold data in disk or Flash-based swap space solves the capacity issue but imposes an unacceptable latency overhead when the cold data is accessed later.

Tiered memory architectures offer an attractive solution in
solving memory inefficiencies with \emph{near} and \emph{far} memory tiers~\cite{tpp}.
Near memory tiers are typically DRAM-based and are low latency, costly, and low-capacity memory tiers. In contrast, far memory tiers are high latency (higher than DRAM, but less than disk or Flash latency), cost-effective, and high-capacity memory tiers. Example far memory tiers include CXL-attached memories~\cite{cxl-website}, non-volatile memories (NVM)~\cite{nvm_paper}, compressed memory pools~\cite{far-memory}, and disaggregated remote memory~\cite{diagg_mem}.  The key idea is to store the frequently used hot data in the near memory tier and cold data in the far memory tier. 

\begin{figure}
    \centering
    \includegraphics[width=\linewidth]{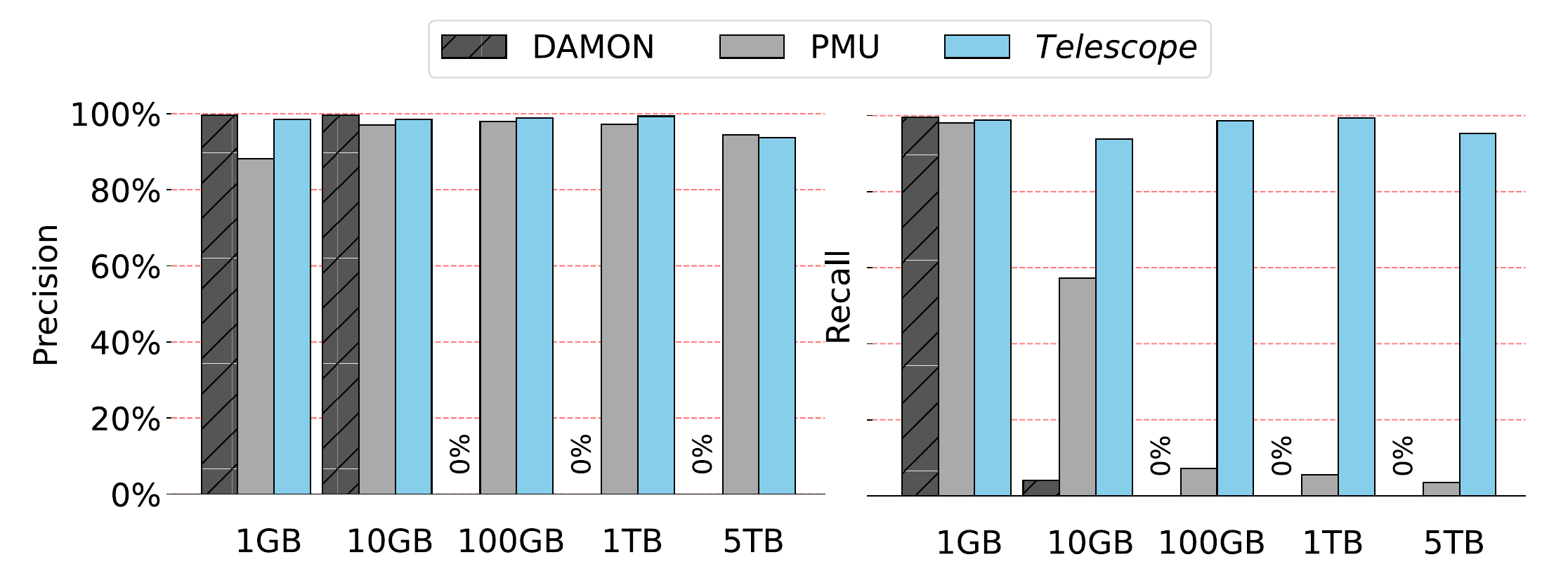}
    \caption{Telemetry efficiency based on precision and recall (\S\ref{subsec:pr-rc}). For \sota techniques, efficiency degrades quickly as the application footprint escalates.}
    \label{fig:teaser}
\end{figure}

However, \textbf{\emph{memory tiering is only as good as the telemetry (hot and cold data detection)}}. Because effective use of tiered memory requires precise and timely identification of hot and cold data sets
and then proactively placing them in appropriate tiers. 
Incorrect or delayed telemetry may place a hot data set in the far memory tier and a cold data set in the near memory tier, resulting in significant performance degradation, which can offset the TCO savings achieved through memory tiering.
\emph{Importantly, AI/ML models that either autotune page placement across memory tiers or predict hot and cold data sets completely depend on precise  telemetry data for offline analysis and training~\cite{far-memory}}.

Unfortunately, existing state-of-the-art telemetry techniques for hot and cold data detection, even though effective at gigabyte scale, are either ineffective or completely fail at terabyte scale. 
For example, techniques that linearly scan the virtual address space~\cite{kstaled,idle-page-tracking} of applications to find hot and cold data cannot provide timely telemetry due to the large number of pages that need to be scanned. Hardware and software-based approaches~\cite{damon,hmkeeper,thermostat,hemem,tpp} that sample accesses to data pages
for telemetry do not scale when the application's working set grows beyond gigabyte scale, as shown in Figure~\ref{fig:teaser}. 

In this paper, we propose \textit{\ptp}, a novel technique for fast and efficient identification of hot and cold data regions.
We observe that different levels of a multi-level page table tree, from the leaf entry to the root entry, have access bits that are updated during a hardware page table walk. This implies that a hot data region will also have hot entries at all levels of a page table tree corresponding to the path of the page table walk. Similarly, if the access bit at a particular level of a page table tree is not set, then none of the data pages under its subtree have recently been accessed and hence are cold. 
We leverage this insight to quickly converge from the root of the page table tree to the actual hot data region and thus efficiently identify gigabytes of cold data regions at terabyte scale.

Since we exploit the natural layout of the page table structure to identify hot and cold data regions, our technique can seamlessly scale beyond terabyte-scale to even petabyte-scale applications on a five-level page table without any additional significant performance overheads.

We implement \ptp in Linux kernel and \texttt{x86\_64} architecture, but \emph{\ptp is portable across hardware architectures} 
that support radix page tables
~\cite{arm-page-table, riscv-page-table, power10-page-table}. 
For a 5\,TB microbenchmark, \ptp achieves 90\%$+$ precision and recall compared to 0\% by Linux kernel's \damon and less than 10\% by hardware counters. For 1--2\,TB memory footprint real-world in-memory database benchmarks, memory tiering based on \ptp
results in 5.6\% to 34\% throughput improvement while the throughput improves marginally for hardware counters and drops for \damon. 

The primary contributions of this paper are as follows:
\begin{itemize}
    \item Compare and contrast the efficiency of the \sota telemetry techniques for terabyte-scale workloads.


    \item Propose \ptp, a fast, efficient, and scalable telemetry technique that can seamlessly scale beyond terabyte scale.

    \item  To the best of our knowledge, \ptp is the first technique to profile page table tree for efficient telemetry.

\end{itemize}

\section{Background}
\label{sec:background}
Before describing the contributions of this paper, we provide the necessary background on modern page table layouts and memory tiering.


\subsection{Page Table}
\label{sec:page_table}

\begin{figure}
    \centering
    \includegraphics[width=.8\linewidth]{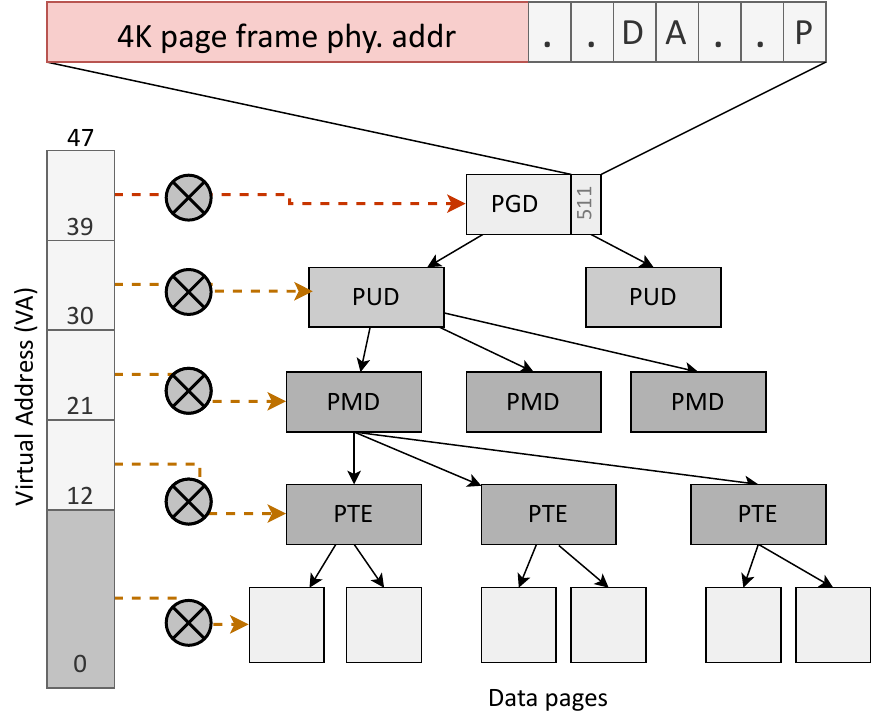}
    \caption{A 4-level page table structure.}
    \label{fig:page_table}
\end{figure}

The page table of a process is a hardware-defined radix tree-based structure maintained by the operating system (OS) to manage the virtual address (VA) to physical address (PA) mappings. A radix page table tree is supported in many hardware architectures such as x86\_64~\cite{intel-docs}, ARM~\cite{arm-page-table}, \mbox{RISC-V}~\cite{riscv-page-table}, \mbox{POWER}~\cite{power10-page-table}.

Modern architectures such as \texttt{x86\_64} support both 4-level and 5-level page table structures. We use the 4-level page table layout, capable of supporting virtual address sizes up to 256\,TB, for our discussion. 
The page global directory (PGD) is the root of the page table tree having 512 entries of size 8 bytes. Each PGD entry 
points to the base physical address of a page upper directory (PUD) and also
maintains a set of flags (e.g., page is \texttt{PRESENT}, page is \texttt{READ\_ONLY}, page has been recently \texttt{ACCESSED}, or page is \texttt{DIRTY}~\cite{intel-docs}).
A similar layout is followed for PUD entries,  page middle directory
(PMD) entries and page table entries (PTE) but at different levels (Figure~\ref{fig:page_table}). 

Upon a TLB miss, the hardware page table walker walks the page table to find the VA to PA mapping. 
During the page table walk, the first 9 bits of the 48 bits VA is used as an index into PGD
to extract the physical address of the PUD. The next 9 bits in the VA index into the PUD to extract the physical address of PMD and  similarly the next 9 bits index into the PMD to extract the physical address of the PTE. Finally, the last 9 bits in the VA is indexed into PTE to extract the physical address of the data page. The remaining 12 bits in the VA are used as an offset in the data page. 

During a page table walk, the hardware sets the \texttt{ACCESSED} bit at all
levels (from PGD to PTE) of the page table tree. 

\subsection{Tiered Memory Architectures}
\label{sec:tiered_memory_architecture}

We provide a brief background on the currently available technologies that can be used as a far memory tier. 

\smallskip
\noindent \textbf{CXL-attached memories.}
 Compute Express Link (CXL)~\cite{cxl-website, cxl, cxl-pooling} enables memory expansion by directly plugging a memory expander card into a server~\cite{cxl-expansion}. The expanded memory serves as a far memory tier for data-intensive workloads~\cite{cxl-expansion-databases,cxl-dl}.

\smallskip
\noindent \textbf{Non-Volatile memories (NVM).}
NVM-based byte-addressable 
memory such as Intel's Optane DC PMM ~\cite{optane, dcpmm} is a high capacity and low bandwidth memory typically used as a far memory tier.
Optane is DDR4 socket compatible and can be plugged into the standard DIMM slots to expand the physical memory to terabyte scale. 

\smallskip
\noindent \textbf{Compressed memory pools.} 
Infrequently accessed pages or cold data pages are compressed and placed in a compressed memory pool such as ZSWAP to reduce memory TCO by reducing the amount of DRAM provisioned on a system~\cite{far-memory}.

\smallskip
\noindent \textbf{Disaggregated remote memory.}
The far memory tier can be provisioned as a remote memory pool which is accessed by the host directly over the network. Accessing a page from a remote memory pool is costly as it has to go over the network to fetch the data~\cite{calciu2021rethinking}
\section{Related Work \& Motivation}
\label{sec:motivation}
\label{sec:related_work}

In this section, we discuss the related works
and highlight their design limitations.
We then motivate the need for novel telemetry techniques that are precise and efficient for terabyte-scale applications and beyond.


Several memory management systems have been proposed for tiered memory systems in recent years~\cite{pond, autotiering, thermostat, kleio, heteroos, far-memory, nimble, radiant}. Several prior works include techniques for hot and cold data identification along with data migration policies and optimizations to enable proactive data placement in near and far memory tiers.
We classify these techniques into three broad groups: \one \textit{linear scanning}, \two \textit{region-based sampling}, and \three \textit{hardware counters} (Table~\ref{tab:related_work}).

\begin{table}
  \centering
  \footnotesize
  \caption{Table summarizing profiling precision (PR) and recall (RC) (\S\ref{subsec:pr-rc}) at gigabyte (GB) and terabyte (TB) scale.}
  \label{tab:related_work}
  {%
 \begin{tabular}   {|L{1.13cm}|L{2.6cm}|c|c|c|c|}
    \hline
     \multirow[c]{2}{*}{\textbf{Group}} & \multirow[c]{2}{*}{\textbf{Prior arts}}& \multicolumn{2}{|c|}{\textbf{GB Scale}} & \multicolumn{2}{|c|}{\textbf{TB Scale}} \\ \cline{3-6}
        &             &\textbf{PR}&\textbf{RC}  &\textbf{PR}&\textbf{RC} \\
    \hline
    Linear scanning       &  TPP~\cite{tpp}, kstaled~\cite{kstaled}, Idle page tracking~\cite{idle-page-tracking}, MGLRU~\cite{mglru}, HeteroVisor~\cite{heterovisor}, AutoTiering~\cite{autotiering},  & High &  High & Low & Low \\
    \hline
    Region-based Sampling & DAMON~\cite{damon}, HMKeeper~\cite{hmkeeper}                  & High &  High & Low & Low \\
    \hline
    Hardware counters     & Thermostat~\cite{thermostat}, HeMem~\cite{hemem}, TPP-Chameleon~\cite{tpp}    & High &  High & High & Low \\
    \hline
    \rowcolor{LightCyan}
    Page table profiling & \textit{\ptp} & High &  High & High & High \\ \hline
  \end{tabular}  
  }
\end{table}

\subsection{Linear scanning}
\label{sec:linear_scanning}

Techniques in this group linearly scan the entire virtual address space of the application to identify hot and cold data by leveraging the \accessed bit in PTE.
It requires two full scans of the virtual address space to identify accessed data pages. The first scan resets the \accessed bit in the PTE entry for every data page while the second scan checks the entire virtual address space to find the data pages with the \accessed bit set.
A set bit indicates that the page was accessed at least once since the last reset~\cite{vmware-exsi, kstaled}.
It periodically scans the virtual address space and checks for data pages that were accessed during a time window. Using this access information, it classifies the data pages into hot and cold sets.

%

\begin{figure}
  \centering
  \includegraphics[width=.7\linewidth]{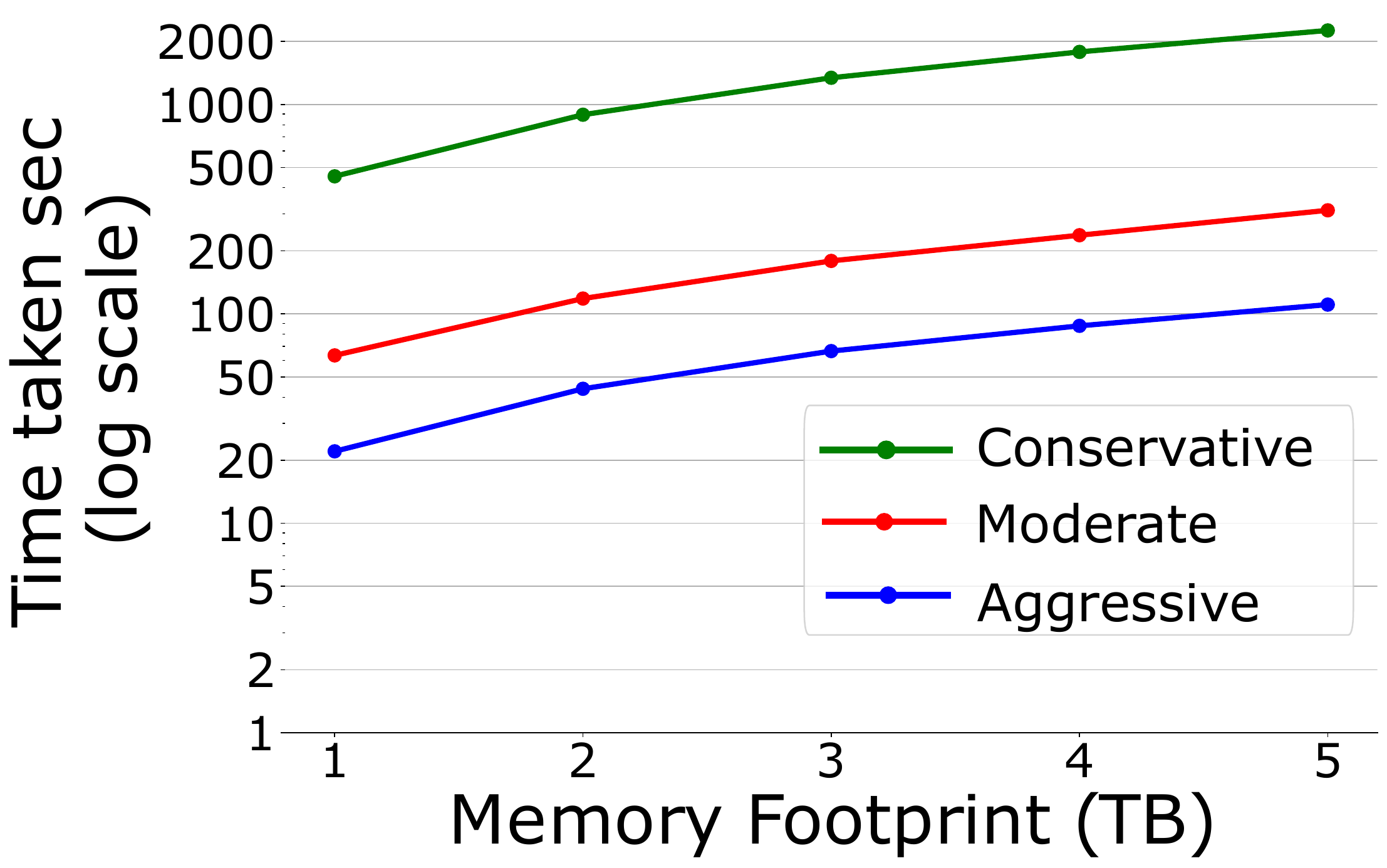}
  \sepblock
  \sepblock
  \footnotesize
  \begin{tabular}{|c|c|c|c|}
\hline
Config   & Aggressive              & Moderate              & Conservative             \\ \hline
CPU util. &  $49.17\% \pm 0.07$ & $19.48\% \pm 1.42$ & $2.78\% \pm 0.43$ \\ \hline
\end{tabular}%
  \caption{Time and compute trade-offs for one full linear scan for workloads with different memory footprints along with the associated CPU overheads.}
  \label{fig:time_taken}
\end{figure}

\sepblock
\textbf{Limitations:} Linear scanning does not scale for workloads with terabytes of memory. The time and compute overheads associated with scanning the application's virtual address space increase with the application's memory footprint.

HeMem~\cite{hemem} points to linear scanning inefficiencies at terabytes scale but misses out the point that linear scanning is a trade off between CPU overhead and scan time. 
We analyze this trade off by implementing a kernel thread in Linux (system details in \S\ref{sec:evaluation}) that 
yields the CPU by sleeping for a fixed duration of time (\textit{conservative}: 100\,ms sleep, \textit{moderate}: 10\,ms and \textit{aggressive}: 0 or no sleep) after flipping the PTE \accessed bits for 256\,MB of data pages scanned. 


As shown in Figure~\ref{fig:time_taken}, for a 5\,TB workload, linear scanning in aggressive mode took over 110 seconds to complete a single scan, but at the cost of significantly high single CPU utilization of 49.2\%. In moderate mode, the CPU utilization drops to 19.5\%, but the scan time increases to 5.2 minutes. But in conservative mode, CPU utilization further drops to 2.8\%, but requires over 37 minutes to complete a single scan.


Note that multiple scans are required to build a precise profile of hot data pages. 
Furthermore, as an application's hot and cold data set can dynamically change over time, scanning cannot be paused. Always-on aggressive scanning results in prohibitively high CPU overheads impacting system and application performance. 
However, employing conservative scanning reduces CPU overheads but increases the scan time and can hence fail to quickly recognize changing data access patterns of applications. This can cause hot data regions to reside in far memory for an extended duration of time. Thus, linear scanning is not suitable for terabyte-scale applications.

\subsection{Region-based sampling}
\label{sec:region_based_sampling}

To reduce the overheads of linear scanning, region-based sampling~\cite{damon} limits the number of data pages that need to be tracked. It divides the application's virtual address space into fixed-size regions to reduce the number of pages it has to track. It then randomly samples one or more data pages in that region and tracks accesses to them using the \accessed bit in PTE. The total number of accesses detected in a given region is assumed to represent the hotness or coldness of the entire region.

A hot memory region thus identified is gradually split into smaller regions to monitor memory accesses at a finer granularity. Adjacent cold regions are merged to form a bigger region to reduce the monitoring overhead. 
DAMON (Data Access Monitor) is one such technique that has been incorporated into the mainline Linux Kernel. 

\sepblock
\textbf{Limitations: }
Although effective for workloads with gigabytes of memory footprint, the method does not scale for terabytes-scale applications.
As the memory footprint increases, the probability of sampling an address belonging to the hot data regions reduces. 
%
In this technique, the convergence to the correct hot data region completely depends on whether the pages belonging to the hot data region is picked by random sampling. If they are not picked for sampling, this technique fails to converge to hot regions.
Increasing the sampling rate increases the probability of finding a hot data region but with increased CPU overheads.

Figure~\ref{fig:teaser} clearly shows that 
region-based sampling is not suitable for 
terabyte-scale applications as the efficiency of hot data detected by \damon deteriorates with the increase in memory footprint.

\subsection{Hardware Counters}
\label{sec:hardware_counters}

Techniques in this group leverage the hardware counters or performance monitoring units (PMUs) to identify an application's hot and cold data pages. PMU events such as retired load/store instructions, TLB misses or L3 cache misses are typically monitored for hotness tracking.
Once a PMU event is enabled for monitoring, the hardware increments a counter at each occurrence of the event, and when the counter overflows, the hardware generates an exception. OS handles the exception and saves the event state to in-memory buffers. The virtual addresses that caused the PMU event are also saved in the buffers, which are then used to identify the hot data set.

\sepblock
\textbf{Limitations:}
Hardware counters are also based on sampling, where PMUs sample the hardware events. 
Similar to region-based sampling, the efficiency of hot data detected by PMU deteriorates with the increase in memory footprint as shown in Figure\ref{fig:teaser} with Intel’s PEBS~\cite{pebs}. 
Increasing the sampling rates improves the probability of hot and cold region detection but can negatively impact the application performance. Because higher sampling rates result in frequent PMU interrupts to the OS. 

In addition, the overheads of this technique are proportional to the size of the hot region. 
For instance, monitoring a 1\,TB hot region using TLB miss event requires generating \textit{at least} 268 million events (one event per 4\,KB page) to  precisely identify the entire hot region. This can generate thousands of PMU interrupts to the OS impacting system and application performance.  
Furthermore, operating systems such as Linux monitor the rate at which PMU interrupts are triggered and automatically lower the sampling frequency if the percentage of time spent in interrupt processing exceeds a certain threshold~\cite{kernel_pmu_events}. 
This automatically reduces the number of samples generated which in turn reduces the precision at which hot data regions are identified.
Hence, hardware counters are also not suitable for terabyte-scale applications.

\subsection{Discussion}
\label{motivation:dis}
The use of 2\,MB huge pages for linear scanning reduces the overheads of linear scanning by an order of magnitude as a single huge page covers 512 base pages of size 4\,KB. 
However, as memory footprint scales to several terabytes, linear scanning at huge page granularity still requires scanning several million huge pages and hence results in high scanning time as observed in HeMem~\cite{hemem}. 
Similarly, for region-based sampling, the probability of sampling hot huge pages can still be low for applications with several terabytes footprint. 
For hardware counters, using huge pages does not improve profiling efficiency as monitoring retired load/store instructions or L3 cache misses to identify hot data set is neutral to the page size used by the application.

Hence, the use of huge pages does not fundamentally solve the telemetry inefficiencies of the state-of-the-art profiling techniques at terabyte scale.


\subsection{Summary}
It can be concluded that linear scanning, region-based sampling, and hardware counters do not enable a fast and efficient identification of hot and cold data sets. Further, their effectiveness degrades quickly as application footprints escalate (Figure~\ref{fig:teaser}).
As memory tiering is only as good as the telemetry, we strongly argue for the need for novel telemetry techniques that are precise, timely, and efficient for \textit{gargantuan} memory footprint applications. 


%




\section{Design Principles}
\label{sec:concept}

In this section, we explain the principles that guide the design of \ptp.

Existing \sota telemetry techniques~\cite{damon, kstaled, vmware-exsi} rely on checking \accessed bits only at the leaf level of the page table tree. 
However, for the past several decades, hardware architectures have supported \accessed bits at all the levels of a page table tree by updating them during the page table walk.
To the best of our knowledge, \ptp is the first technique that exploits this hardware feature by dynamically profiling different levels of a multi-level page table tree to precisely and efficiently identify hot and cold data regions at terabyte scale.

\ptp leverages the following key insight: 
as \accessed bits at all levels of the page table tree 
are updated during a hardware page table walk, a hot data page should also have a hot PMD, PUD, and PGD entry for a memory and TLB-intensive application. Similarly, if the access bit in a PGD entry (or a PUD/PMD entry) is not set, then none of the memory regions represented by the PGD entry (or PUD/PMD entry) subtree are accessed and hence can be considered as cold.

\ptp profiles \accessed bits at the higher levels of
the page table to initially  identify hot regions at coarser granularity as they cover larger virtual address mappings. Upon detecting accesses, \ptp dynamically profiles lower levels of the page table tree to converge to hot regions.

\begin{figure}
\centering
\includegraphics[width=.9\linewidth]{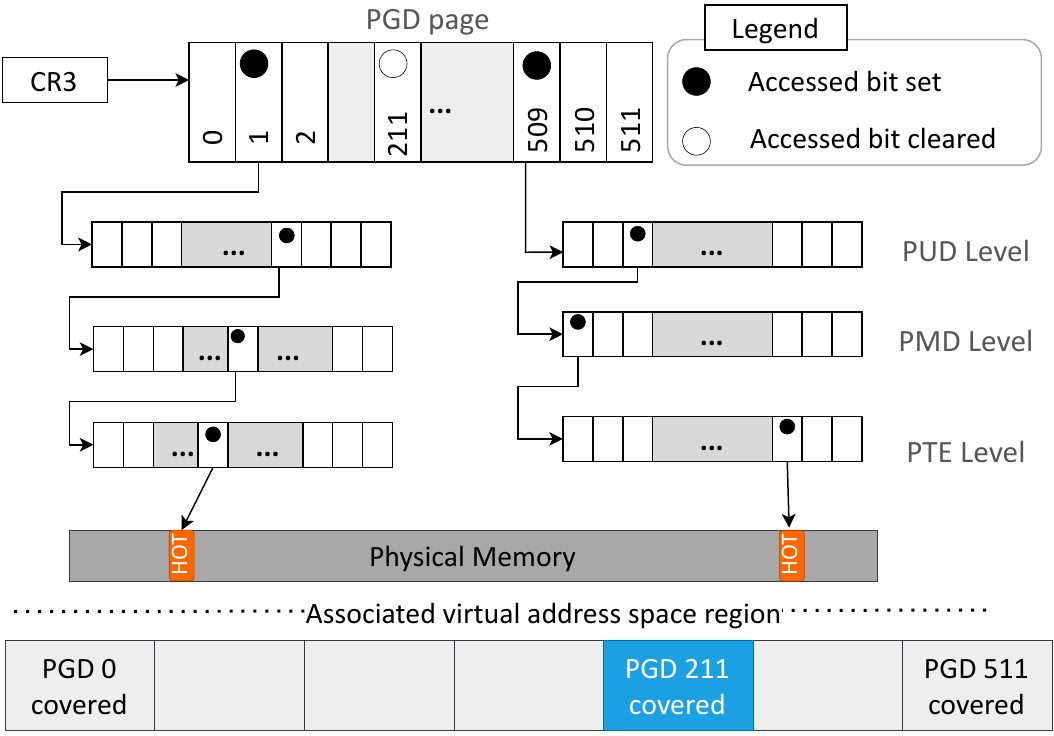}
\caption{Diagram depicting the high-level overview of \ptp}
\label{fig:pt_profile_concept}
\end{figure}

We explain \ptp with an example. Consider a terabyte-scale application with hot data pages as shown in Figure~\ref{fig:pt_profile_concept}.
To identify a hot data page, \ptp starts profiling at the PGD level by resetting the access bits and periodically checking if they are set by the hardware page walker. 
As a single PGD entry covers 512\,GB of virtual to physical address mapping, if the access bit for a PGD entry is set then one or more data pages in the 512\,GB PGD subtree is hot. 

As PGD entries 1 and 509 have the \accessed bit set (Figure~\ref{fig:pt_profile_concept}), \ptp dynamically traverses down the page table tree corresponding to these PGD entries to profile at PUD level. 
\ptp resets the \accessed bit at the PUD level and periodically checks if they are set by the hardware page walker.
If the access bit for a PUD entry is set then one or more data pages in the 1\,GB PUD subtree is hot (each PUD entry covers 1\,GB mapping). 
Similarly \ptp traverses down the page table tree by dynamically profiling at PMD and PTE levels if the \accessed bits are set to find the actual hot data page. 
As \ptp traverses down from PGD to PTE it converges from a large 512GB region to the actual 4K hot data page. 

Now consider a scenario where  
the \accessed bit for a PGD entry that was cleared during profiling is still not set (PGD entry 211 in Figure~\ref{fig:pt_profile_concept}). 
In such a case the entire 512\,GB virtual address space subtree corresponding to PGD entry 211 is cold; it is not required to traverse down the page table tree any further. 
This way
several gigabytes of cold regions can be quickly identified without enumerating individual data pages. 

\sepblock
\noindent \textbf{Summary.} \ptp at every iteration converges to the hot data set by traversing down the tree (similar to search technique
in a tree data structure) to a set of subtrees that
contain hot data pages (i.e., for entries with \accessed bit set at that page table level) while it stops further traversing down the subtree if the \accessed bits are not set to identify the cold data pages.

\section{\ptp Design}
\label{sec:design}



\begin{figure*}
\centering
    \subfloat[\label{fig:pt-level-bounded1}]{%
        \includegraphics[width=0.47\textwidth]{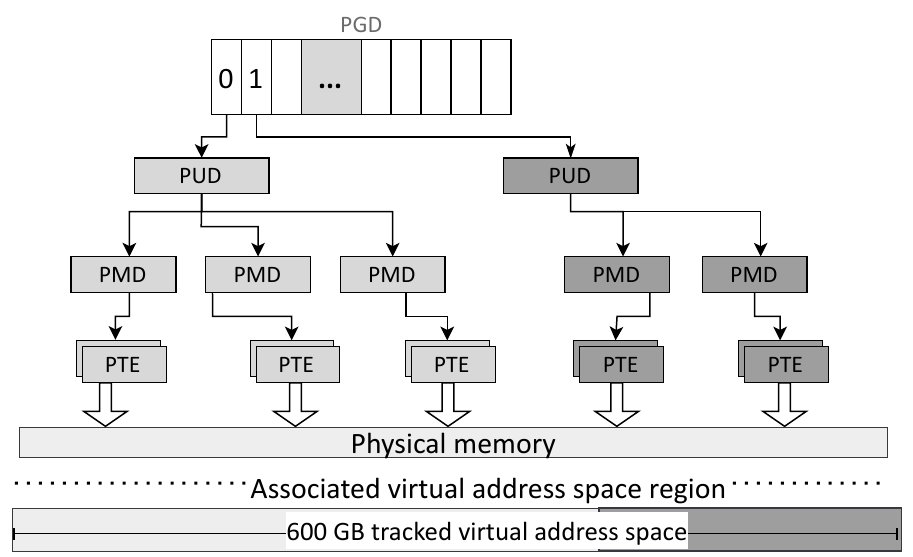}
    }
    \hfill
    \subfloat[\label{fig:pt-level-bounded2}]{%
        \includegraphics[width=0.43\textwidth]{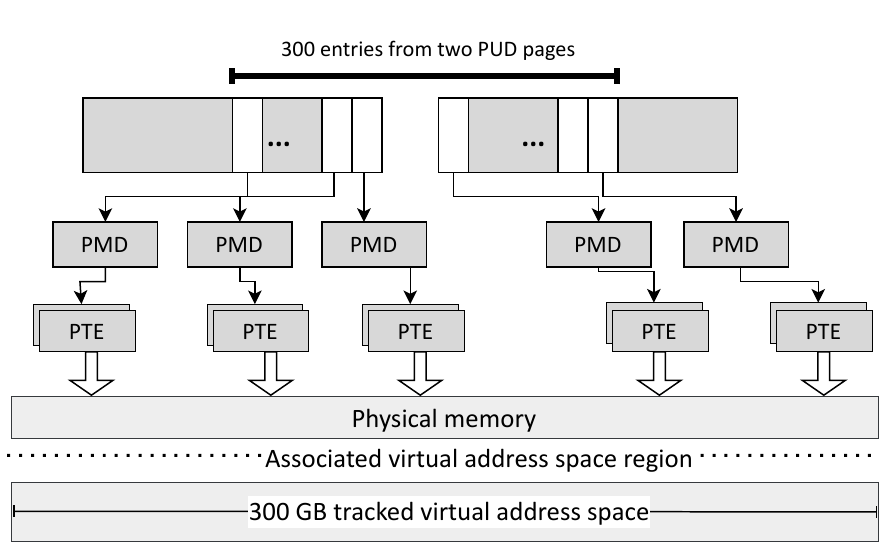}
    }
    \caption{Bounded variant used by \ptp to select page table level at which to track the accessed bits}
    \label{fig:pt-level-bounded-all}
\end{figure*}

\begin{figure}
\centering
\includegraphics[width=\linewidth]{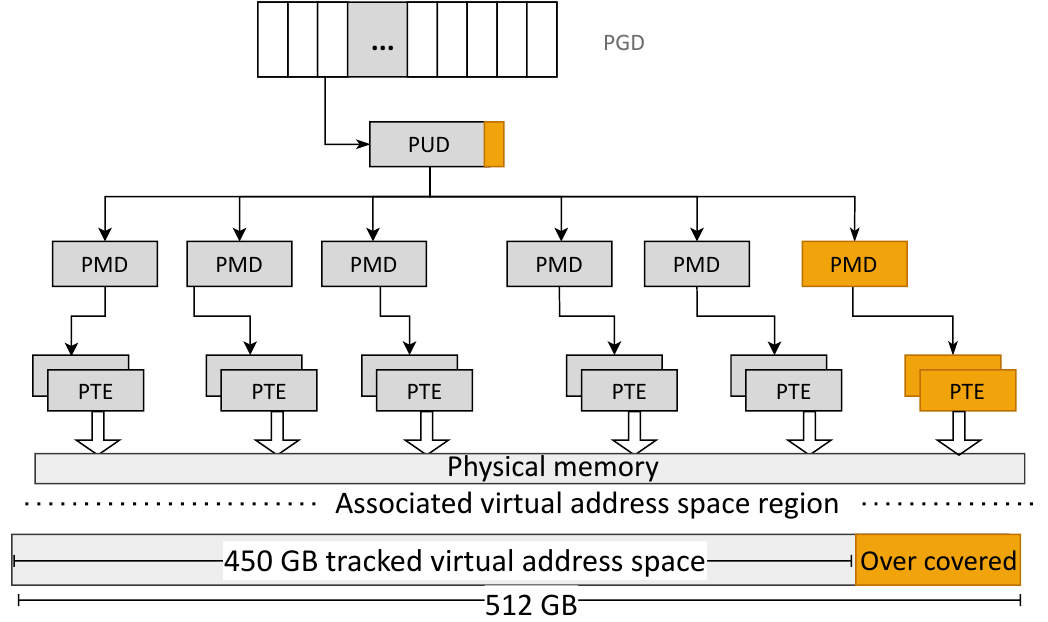}
\caption{Flex variant used by \ptp to select page table level at which to track the accessed bits}
\label{fig:pt-level-overbounded}
\end{figure}

\ptp introduces a novel technique to identify hot and cold data pages in a workload's memory footprint using page table profiling.
In this section, we explain the design of \ptp in detail.

\ptp has two main components (i) region management and (ii) region profiling as explained below.  

\subsection{Region management}

\ptp's region management is inspired by the design employed in the Linux kernel for DAMON~\cite{damon}
as we find it efficient.
\ptp decomposes the workload's virtual address space into a set of equally-sized regions to begin with.
A profiling window is used during which data accesses to each region are monitored.
Based on the number of accesses seen, 
each region is assigned a score that reflects the hotness or coldness of the entire region.
At the end of each profiling window, the following actions are performed: (i) each region is split into random-sized small subregions. We find random splitting of regions
employed in the Linux kernel effective under the dynamically changing memory access patterns of the workloads,
(ii) adjacent regions with similar hotness or coldness score are merged into a bigger region 
and (iii) information is provided to user space regarding the number of regions, the virtual address range of the regions 
and the associated hotness or coldness score. This information can be used  to take suitable actions such as migrating data pages to appropriate tiers or can be fed to AI/ML models for offline training.

Splitting ensures that the regions that contain hot data pages are narrowed down to precision with time,
while merging ensures that cold regions are tracked at a coarser granularity.

\subsection{Region profiling}

Region profiling is the core and critical component of \ptp that precisely identifies hot and cold data regions. 

Each region identified by the region management subsystem is profiled independently in every profiling window.
For each region, in each profiling window, multiple profiling samples are
recorded at regular intervals. 
For each sampling interval, a page table entry at one of the levels from PTE to PGD
is identified for profiling.
The \accessed bit for the identified page table entry is reset at the beginning of the
sampling interval  
and \ptp checks whether the reset \accessed bit is set at the end of the sampling interval.
If set, then one or more data pages covered by the identified page table 
entry were accessed 
during the sampled interval. Hence the access count for the region is incremented.
At the end of the profiling window, \ptp performs the region management actions such as splitting and merging as discussed before for all the regions.

We implement two different profiling variants (i) \textit{bounded} and (ii) \textit{flex},
which identifies a page table level from PTE to PGD for profiling.
The variants offer two different trade-offs between convergence aggression and profiling accuracy.

\subsubsection{Bounded variant}

This variant ensures that for each region, the identified page table entry is
at the highest page table level whose address range is within the region bounds.
That is, the page table level identified should not span multiple regions.
Tracking the accessed bit at the highest possible level of the page table 
increases the likelihood of convergence, as just a single access bit 
can track data accesses for a large virtual address range.

Consider the example in Figure~\ref{fig:pt-level-bounded1} with a region size of
600\,GB. In this example, the highest possible page table level that \ptp can pick is PGD. The virtual address space covered by PGD entry 0 is within the 
region bounds and does not span multiple regions.
By profiling PGD entry 0, \ptp can quickly identify even a single page access in
the 512\,GB virtual address space. 

However, PGD entry 0 does not cover the entire 600\,GB region. Hence, for the subsequent
sampling intervals, \ptp should pick a page table entry that includes the rest of the
88\,GB to ensure coverage of the entire region.
But for the remaining 88\,GB, PGD entry 1 cannot be used as it covers virtual address space
from 512\,GB to 1024\,GB, which is beyond the address space bounds of the region being profiled. Therefore,
the highest possible page table level that \ptp can pick for profiling the 88\,GB portion is PUD. 

As \ptp progresses, the hot regions are split, resulting in smaller regions. 
Now, the virtual address ranges of the higher levels of the page table do not fit within
the region’s bounds, forcing \ptp to pick lower levels. Nevertheless, this facilitates
the precise convergence of regions to actual hot pages within the application’s memory footprint.
For example, consider a region with size 300\,GB. \ptp cannot pick a
PGD entry as it covers virtual address space beyond this region.
In such scenarios, PUD entry is the highest page table level whose address range is within the region bounds. This is also the case when the region is mapped by two PUD pages as shown in Figure~\ref{fig:pt-level-bounded2}. 
Hence, one of the 300 PUD entries is randomly picked for profiling during every sampling interval.

Similarly, the highest page table level that can be picked for profiling can be a PMD or PTE entry depending on the size of the region being profiled.

\subsubsection{Flex variant}
This variant requires that the identified page table entry is
at the highest page table level but the address range of the picked entry need not be always within the region bounds.
\ptp can be flexible and go beyond the region's bounds but with a certain error threshold.
This ensures better region coverage but at the cost of accuracy.  

Consider the example in Figure~\ref{fig:pt-level-overbounded} with a region size of
450\,GB where the error threshold is 15\%. The virtual address space covered by PGD entry 3 
exceeds the region bounds by 72\,GB, but is well within the error threshold. Hence, \ptp
is allowed to pick PGD entry 3 for profiling. 
Consider another example where the region size is 300\,GB. 
\ptp cannot pick a PGD entry as the virtual address space covered by a PGD entry is beyond the error threshold.
In such scenarios, \ptp falls back to the bounded variant where
one of the 300 PUD entries is randomly picked for profiling
during every sampling interval.



\section{Evaluation}
\label{sec:evaluation}

In this section, we compare and contrast \ptp with other state-of-the-art telemetry techniques.
Our evaluation answers the following questions.

\begin{itemize}

    \item How well does \ptp perform vis-a-vis the state-of-the-art in quickly and accurately identifying hot data?(\autoref{subsec:microbenchmarks}).

    
    \item How do the incurred overheads of \ptp compare with other techniques? (\autoref{subsubsec:overheads}).

    \item How well does \ptp perform on real-world big data applications? (\autoref{subsec:real-world}).
    
\end{itemize}

\subsection{Evaluation setup}

We use a tiered memory system with an Intel Xeon Gold 6238M CPU having 4 sockets, 22 cores per socket, and 2-way HT for a total of 176 cores. It has a DRAM-based near memory tier with 768\,GB capacity and a far memory tier with Intel's Optane DC PMM~\cite{optane} configured in flat mode (i.e., as volatile main memory) with 6\,TB capacity for a total of 6.76\,TB physical memory.
We run Fedora 30 and use Linux kernel 5.18.19 for our evaluation. We use 4\,KB pages unless otherwise explicitly mentioned.

\subsubsection{Telemetry techniques}
To evaluate the performance of \ptp 
we pick one representative technique each from region-based sampling and hardware counters.
We do not evaluate linear scanning-based technique due to prohibitively high CPU overheads (48\%$+$) or time taken to complete a single scan (37 mins) at terabyte scale. 

\sepblock
\noindent \textbf{DAMON~\cite{damon}.} \damon is a region-based sampling technique that is part of the Linux kernel.
We use two configurations for DAMON -- \textit{moderate} and
\textit{aggressive}. \textit{Moderate} (MOD) uses the default values of 5\,ms sampling interval and 200\,ms profile window (or aggregation interval) 
thus generating 40 samples per profile window. 
\textit{Aggressive} (AGG) uses 1\,ms sampling interval with 200\,ms profile window generating 200 samples per profile window. \textit{Aggressive} consumes more CPU cycles as it samples more frequently. The rest of the DAMON parameters are set to the Linux kernel default values. 
We do not include results with different profile windows because they provide no additional insights as the trend remains the same.

\sepblock
\noindent \textbf{PMU.} We use Intel PEBS (Processor Event-Based Sampling), a hardware-based performance monitoring unit (PMU) available on Intel processors~\cite{pebs}. 
PEBS can monitor pre-defined hardware events and can capture additional information such as the virtual address that caused the event. We monitor
\texttt{MEM\_INST\_\linebreak[0]RETIRED.\linebreak[0]ALL\_\linebreak[0]LOADS\_\linebreak[0]PS} and 
\texttt{MEM\_INST\_\linebreak[0]RETIRED.\linebreak[0]ALL\_STORES\_\linebreak[0]PS}~\cite{hemem} events that sample all retired load and store instructions. 
We drive PEBS using the \texttt{perf} tool available in Linux.
We evaluate PEBS with two different sampling frequencies: 10 kHz and 5 kHz for \textit{aggressive} and \textit{moderate} configurations, respectively.
Higher the sampling frequency higher the overheads, as PEBS generates frequent interrupts. 

\sepblock
\noindent \textbf{\ptp.}
We evaluate two variants of \ptp: \textit{bounded} and \textit{flex}, which differ in the way a page table level is picked for profiling as explained in detail in the design section (\S\ref{sec:design}).
Both variants of \ptp are configured to use 5\,ms sampling interval and  200\,ms profiling window.
We use different error thresholds at different levels of the page table tree for the \textit{flex} variant. At PUD, the error threshold is kept low at 15\% as it covers a larger region while at PMD and PTE it is set to 25\%.

\subsection{Microbenchmarks}
\label{subsec:microbenchmarks}
To simulate different memory access patterns we use \textit{memory access simulator}, or \masim~\cite{masim}, a widely used utility by the Linux kernel developers~\cite{pro_swap_masim_use, damon_demo_masim_use2, masim_path_use3, masim_damon_test_use4}.
We generate stable access patterns, as page access patterns remain stable for several minutes to hours in production workloads~\cite{tpp}. In addition, we also demonstrate sensitivity of profiling techniques to changes in access patterns. 

We fix a bug we found both in MASIM and DAMON, to support terabyte-scale workloads, by using a 64-bit random value instead of 32-bit to generate accesses to memory regions greater than 4\,GB. We also optimize \masim to perform multi-threaded memory allocation to reduce the initialization time.

\smallskip
\noindent \textbf{Heatmaps.}
We generate heatmaps to visualize the profiling efficiency. The x-axis in the heatmap is the time, and the y-axis is the virtual address offset in the heap of the workload. For example, if the base virtual address of the heap
is \texttt{addr}, then 1\,TB value on the y-axis represents \texttt{addr}$+$1\,TB. The red color represents hot regions, and the rest are cold regions, as reported by the telemetry techniques. 

\smallskip
\label{subsec:pr-rc}
\noindent \textbf{Precision and recall.}
We quantify telemetry capabilities using two key metrics - \textit{precision} and \textit{recall}~\cite{hmkeeper}.
Precision is the ratio of correctly identified hot pages to the total number of identified hot pages, i.e., the
fraction of the memory identified as hot by the telemetry technique which is indeed hot as per the workload's actual access pattern.
Recall is the ratio of correctly identified hot pages to the number of actual hot pages in the workload i.e.,
the fraction of the workload's actual hot pages that was correctly identified as hot. 

To compute precision and recall for DAMON and \ptp, we use the region data as reported during every profile window. PMU counters using PEBS do not report any region data, but report the virtual address of the profiled events. We use a 2\,MB tracking granularity as used in HeMem~\cite{hemem} to ensure that we do not underestimate the hot data regions of the application by tracking at 
finer granularities.



\subsubsection{Multi phase}
        

\begin{figure*} 
    \centering
    \begin{minipage}[t]{0.48\linewidth}
        \centering
         \subfloat[DAMON-MOD\label{fig:damon-mod-stairs}]{%
        \includegraphics[width=0.48\linewidth]{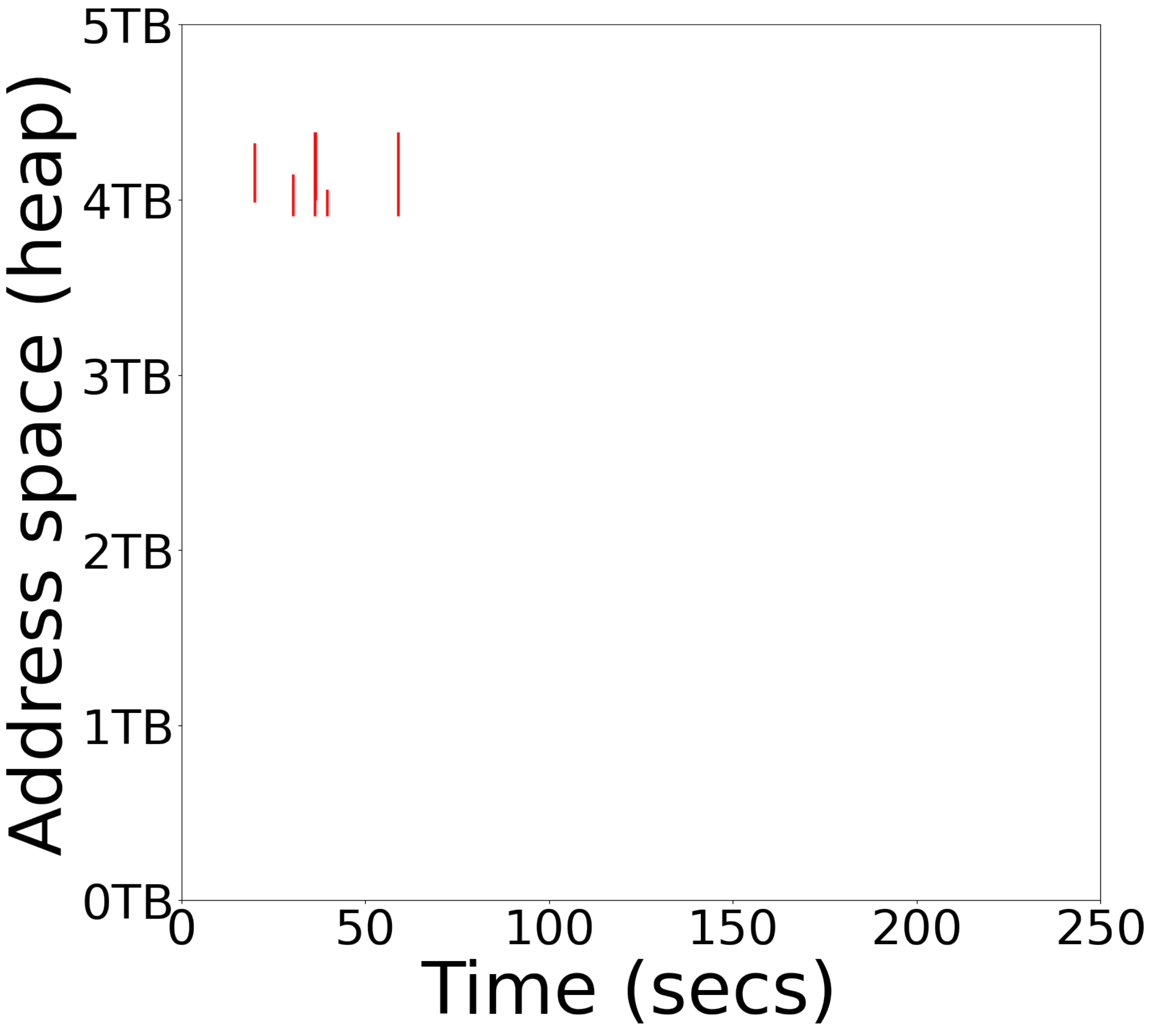}
    }
    \subfloat[DAMON-AGG\label{fig:damon-agg-stairs}]{%
        \includegraphics[width=0.48\linewidth]{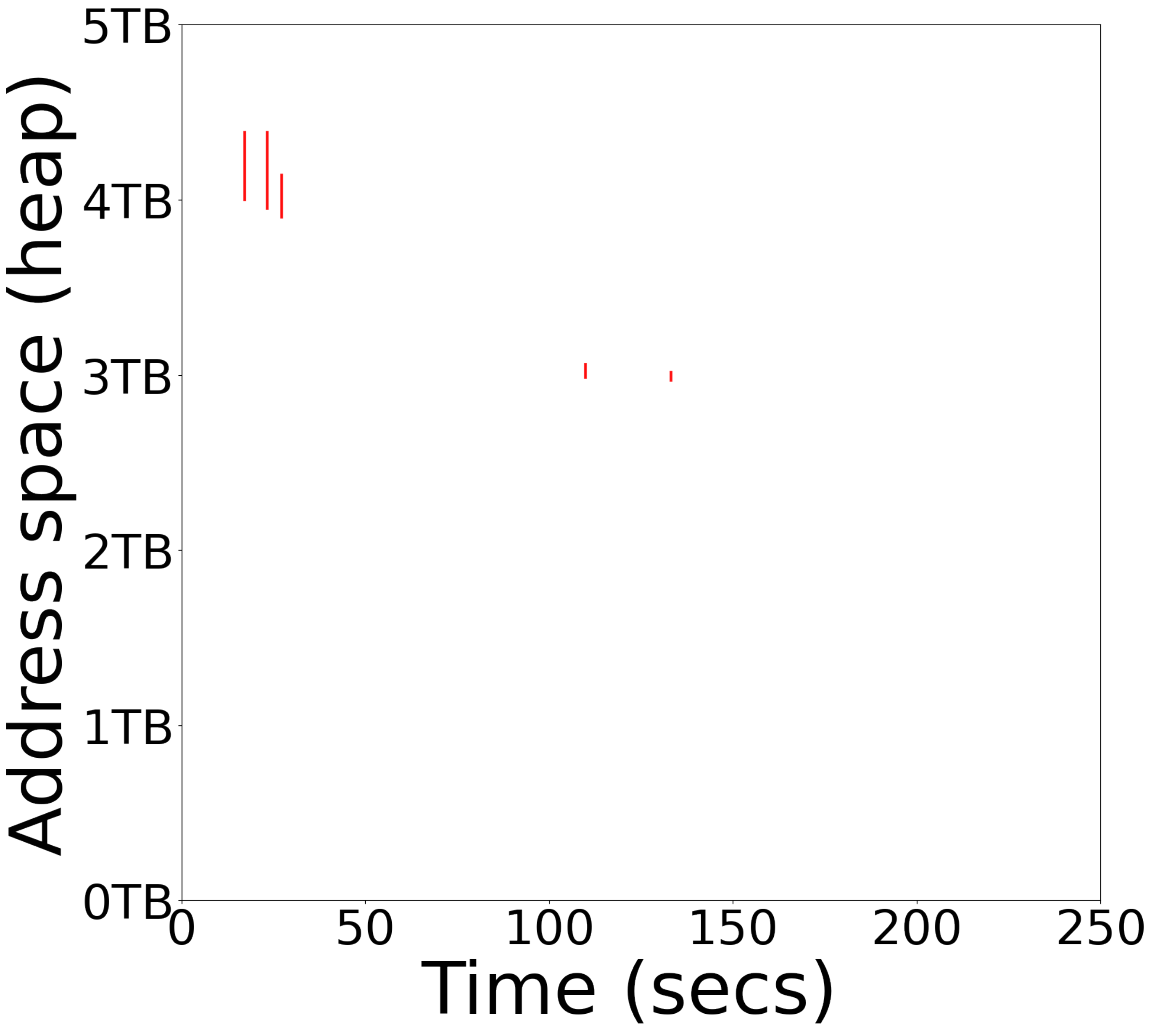}
    }

     \subfloat[PMU-MOD\label{fig:pebs-mod-stairs}]{%
        \includegraphics[width=0.48\linewidth]{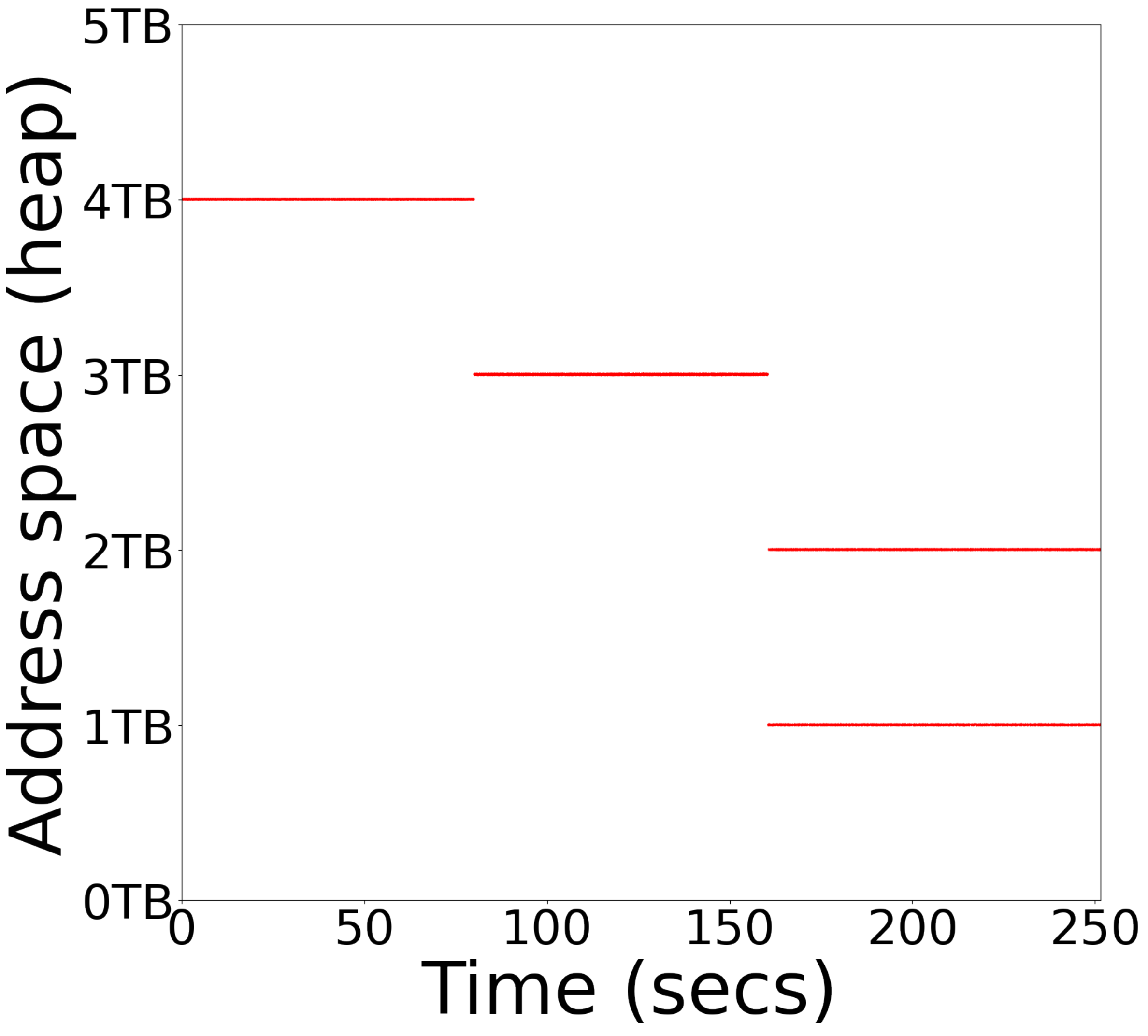}
    }
    \subfloat[PMU-AGG\label{fig:pebs-agg-stairs}]{%
        \includegraphics[width=0.48\linewidth]{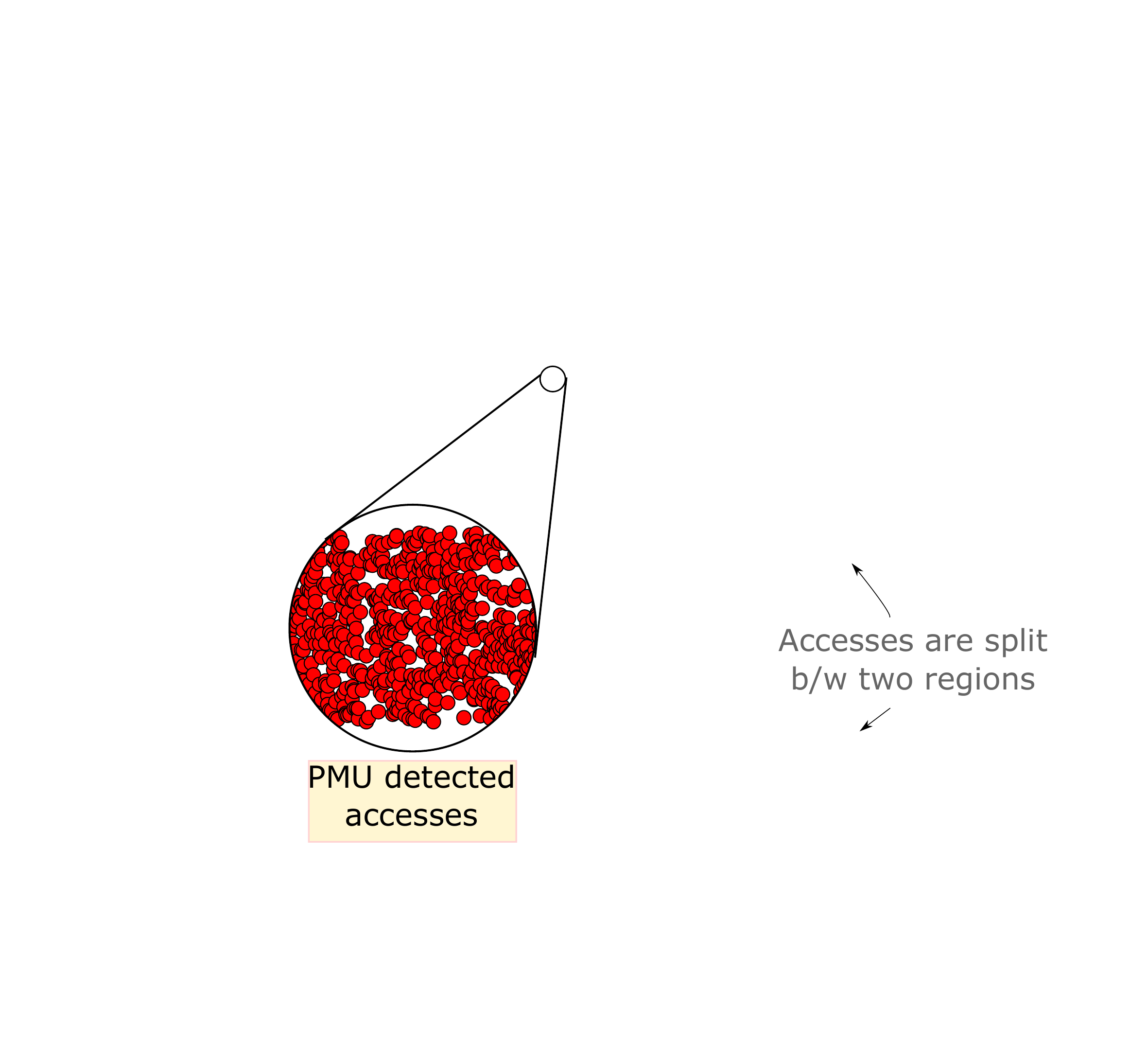}
    }

       \subfloat[\ptp-BND\label{fig:ptp-mod-stairs}]{%
        \includegraphics[width=0.48\linewidth]{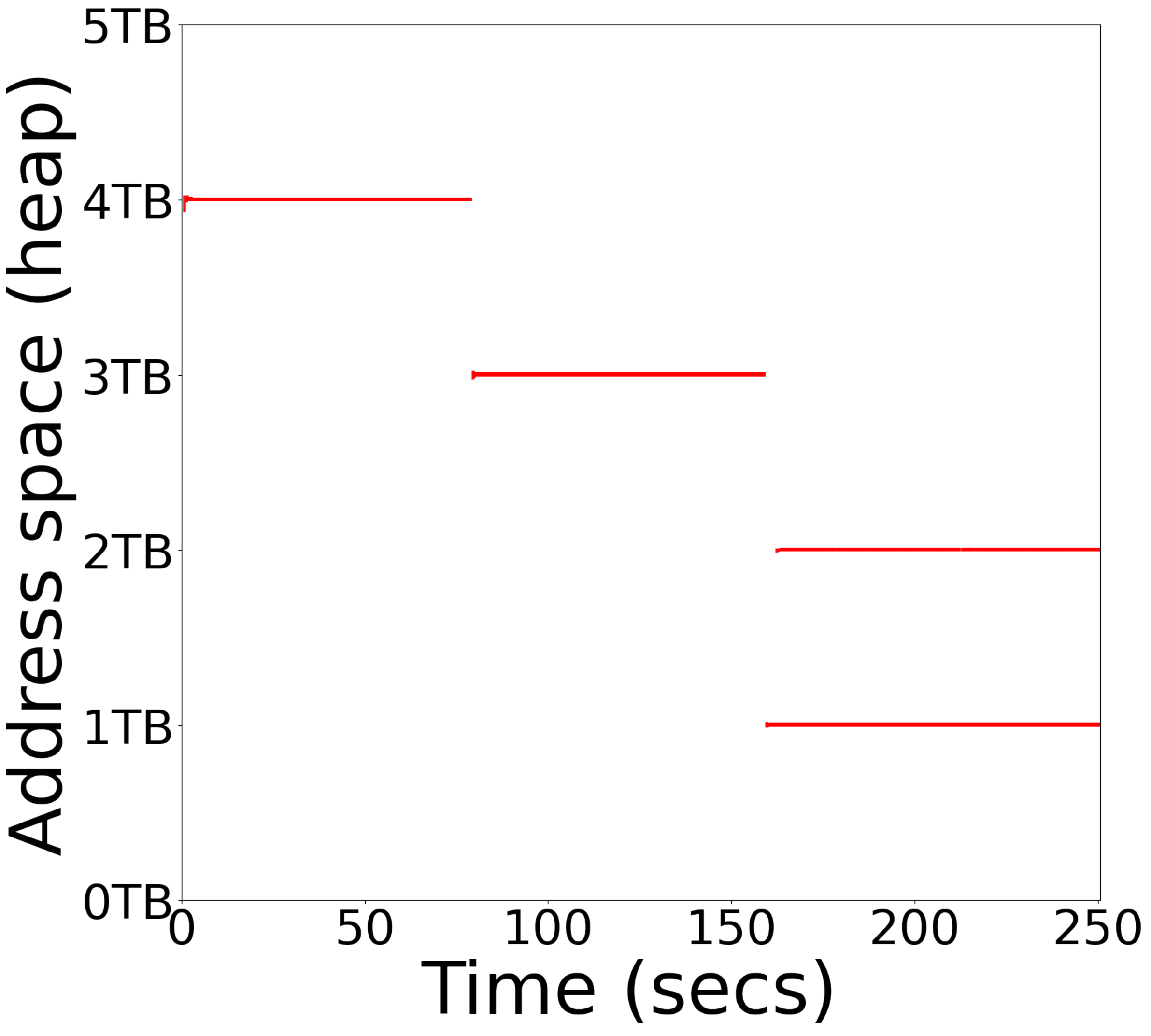}
    }
    \subfloat[\ptp-FLX\label{fig:ptp-agg-stairs}]{%
        \includegraphics[width=0.48\linewidth]{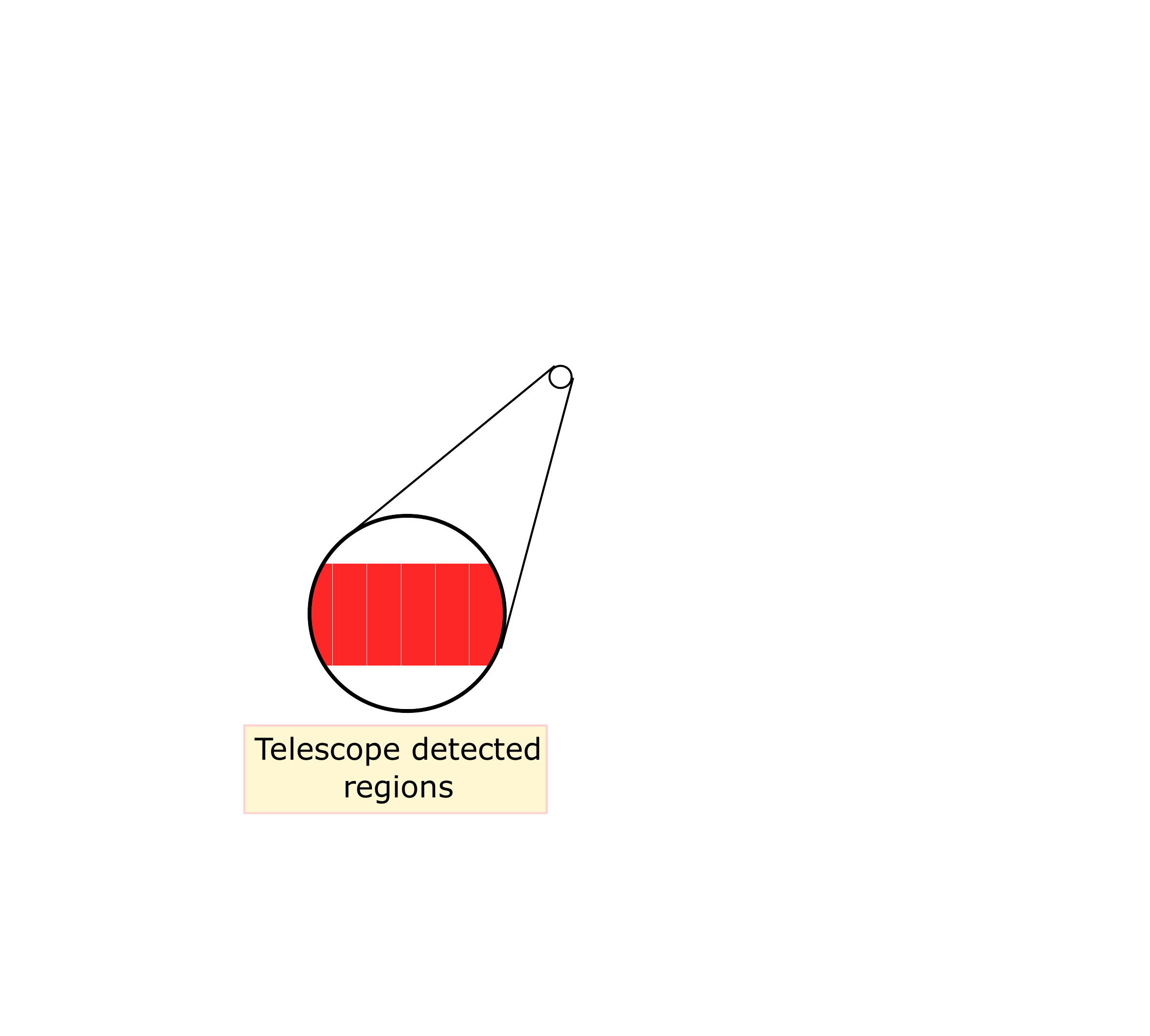}
    }
    
        \caption{Heatmaps for \textit{multi-phase} microbenchmark. The zoomed regions show how the heatmaps differ for region-based \ptp technique and event-based PMU technique.}
        \label{fig:heat_maps}
    \end{minipage}
    \hfill
    \begin{minipage}[t]{0.51\linewidth}
        \centering
             \subfloat[\damon Precision\label{fig:damon_acc}]{%
        \includegraphics[width=0.49\linewidth]{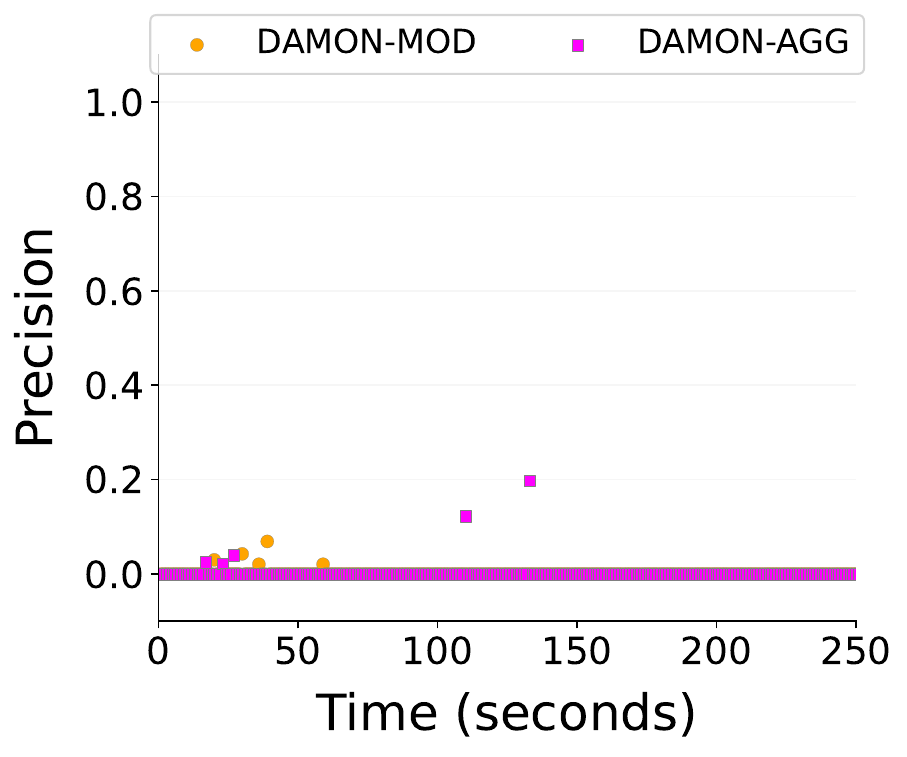}
    }
    \subfloat[\damon  Recall\label{fig:damon_cov}]{%
        \includegraphics[width=0.49\linewidth]{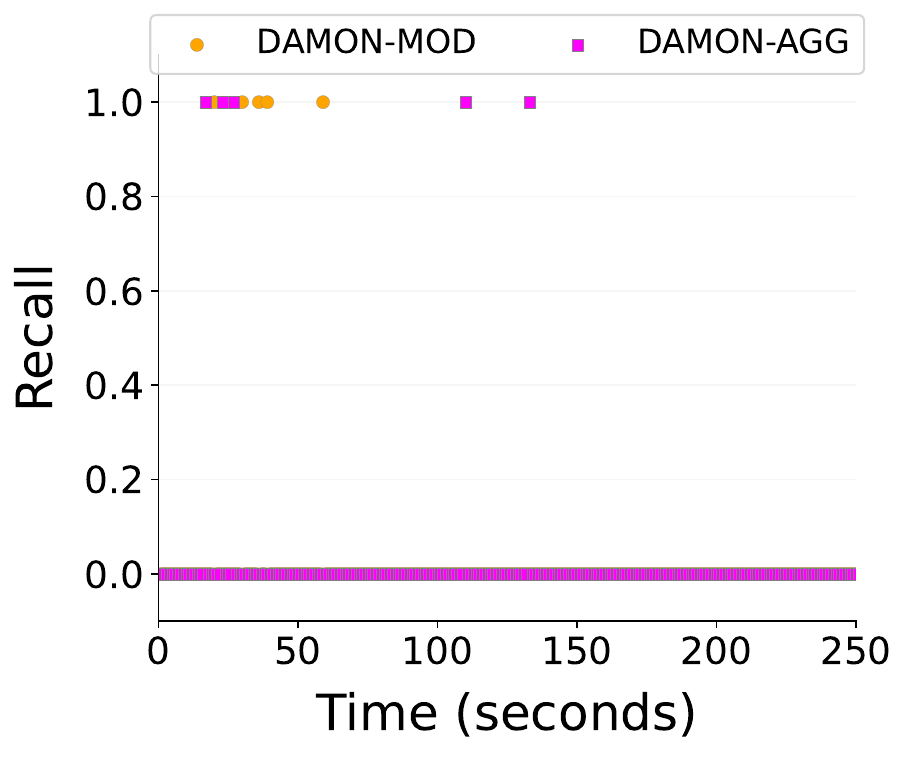}
    }

       \subfloat[PMU Precision\label{fig:pebs_acc}]{%
        \includegraphics[width=0.49\linewidth]{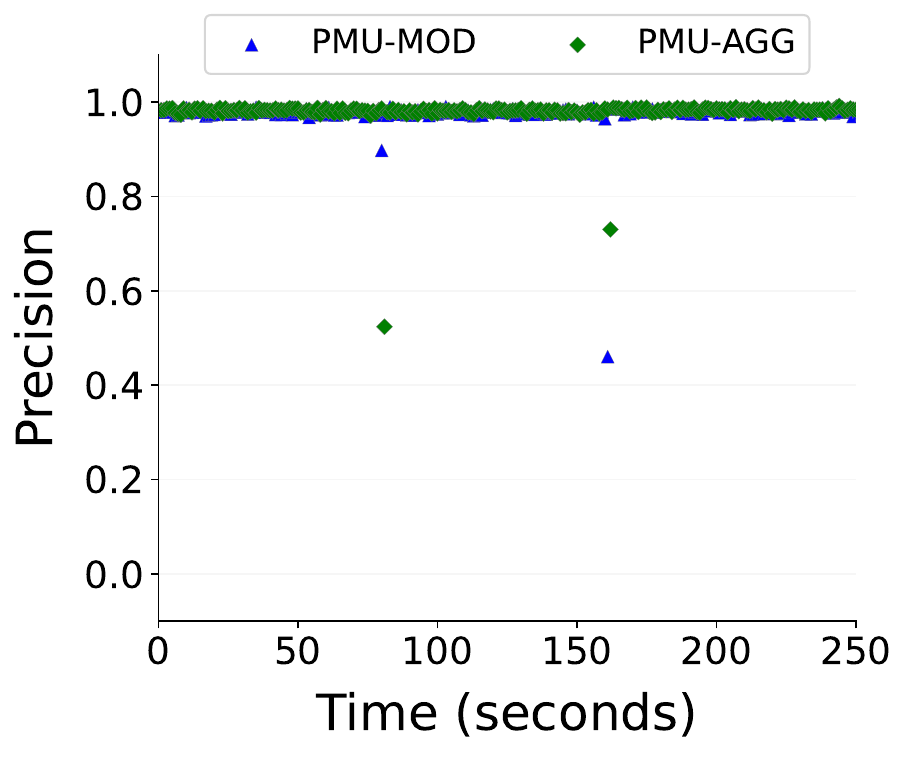}
    }
    \subfloat[PMU Recall\label{fig:pebs_cov}]{%
        \includegraphics[width=0.49\linewidth]{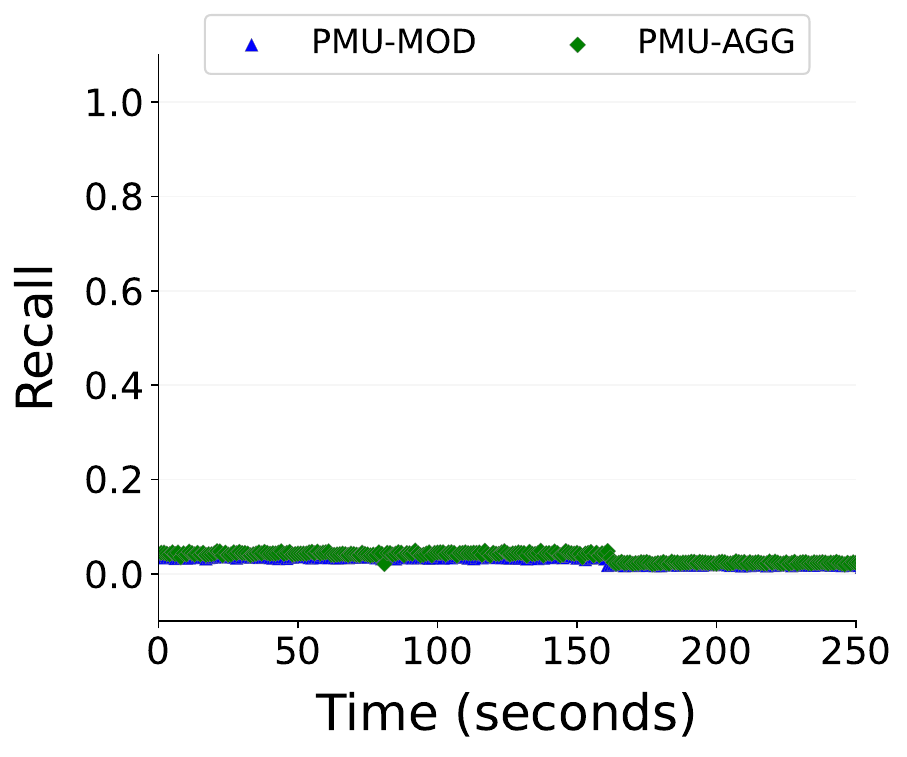}
    }

       \subfloat[\ptp Precision\label{fig:ptp_acc}]{%
        \includegraphics[width=0.49\linewidth]{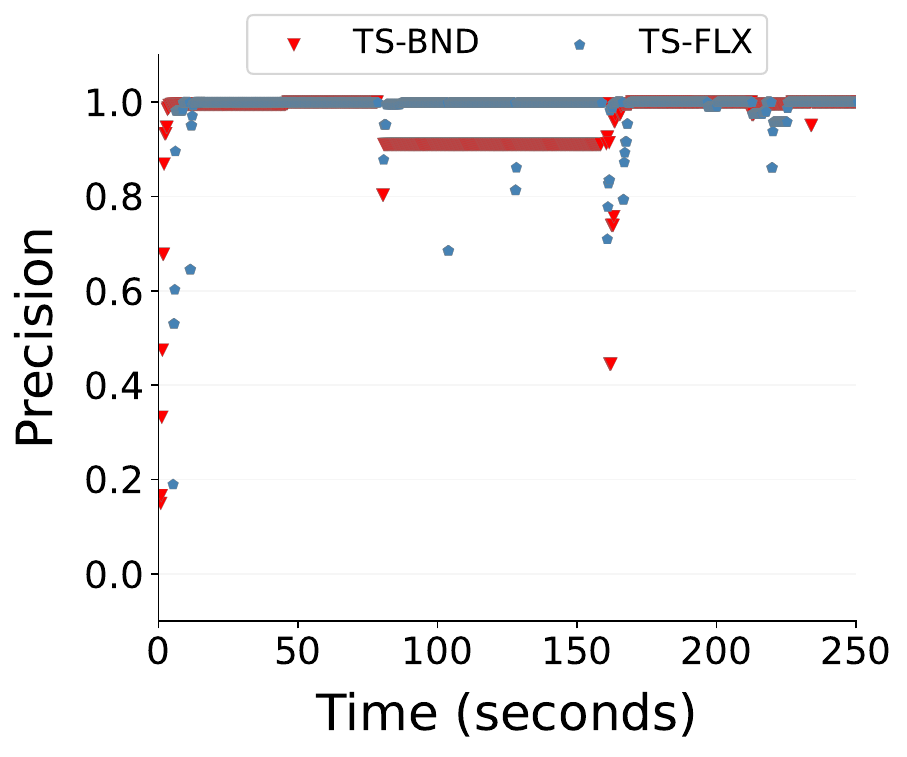}
    }
    \subfloat[\ptp Recall\label{fig:ptp_cov}]{%
        \includegraphics[width=0.49\linewidth]{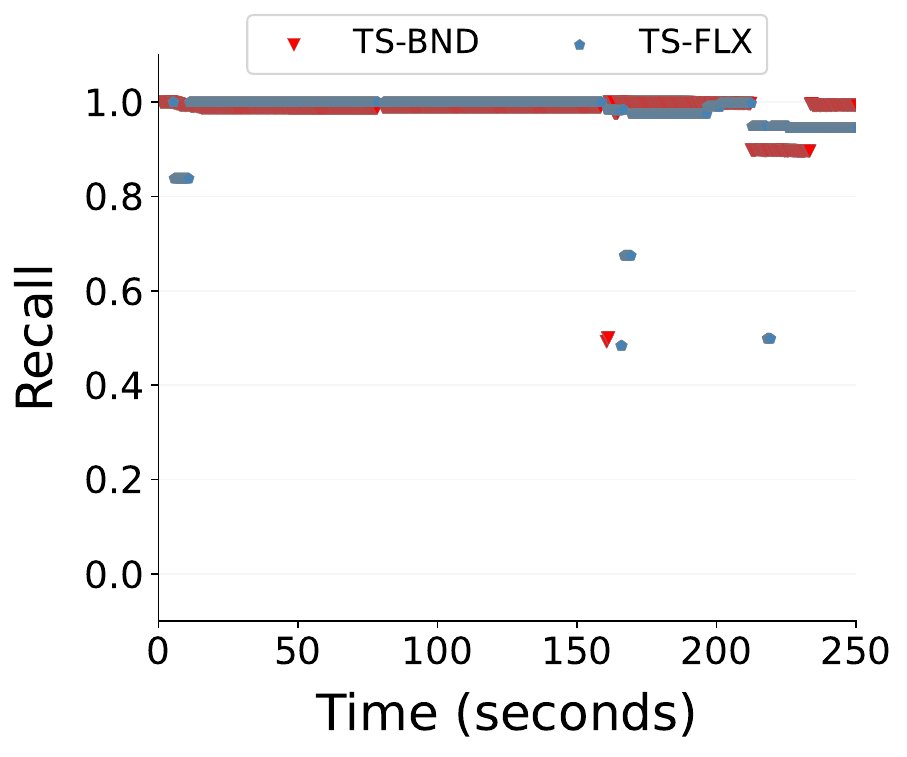}
    }
    
        \caption{Precision and recall values for \textit{multi-phase} microbenchmark}
        \label{fig:precision_recall}
    \end{minipage}

\end{figure*}

The goal is to test the three important hot data identification capabilities: (i) speed and accuracy in identifying hot regions, (ii) sensitivity and responsiveness to dynamically changing hot regions, and (iii) speed and accuracy in identifying multiple hot regions in the entire heap.

We configure MASIM to allocate 5\,TB of heap and simulate access patterns in three different phases to test the capabilities
mentioned above. In the first phase, MASIM performs data loads by randomly picking an address within a 10\,GB region. The second phase is the same as the first phase but on a completely different 10\,GB region. In the third phase, MASIM performs data loads by randomly picking an address from two different 10\,GB regions. Data access patterns in real workloads generally remain stable for minutes to hours~\cite{tpp}, so this microbenchmark is representative of real-world access patterns.

The generated heatmaps are shown in Fig~\ref{fig:heat_maps}.
At the beginning of the first phase \damon
briefly detects few accesses to hot regions, but fails to converge to it in the subsequent profiling windows. Both variants of \damon completely fail to capture hot regions in the second and third phases.
This is because, at the terabyte scale, 
the probability of the sampled address belonging to the hot data regions is low. 
Hardware-based PMU captures the hot regions and can identify hot regions in all three phases. 
\ptp successfully captures the hot regions in all three phases. 

   
    

\smallskip
\noindent \textbf{Precision and recall.} Figure~\ref{fig:precision_recall} shows the precision and recall for the multi-phase microbenchmark. 
\damon's precision and recall are mostly 0 as it fails to detect hot regions.
PMU's precision values are always close to 1 as they include data only for the actual events
(there are no events outside of the hot region). However, the recall for both variants of PMU is less than 0.1 because, covering the entire hot region requires generating millions of events which is not possible (as discussed before in limitations of hardware counters, \S\ref{sec:hardware_counters}).

Both variants of \ptp outperform both DAMON and PMU in all the phases. 
\ptp's precision and recall remain above 0.9 for all three phases of the benchmark. The precision for \ptp momentarily drops during the phase change (at around 80 and 160 seconds) but quickly recovers. 
This clearly demonstrates that only \ptp passes our versatility test.

\smallskip
\noindent \textbf{Huge pages.}
We repeat the experiments with transparent huge pages or 2MB pages enabled to compare and contrast the efficiency of telemetry techniques. 
With huge pages, \damon performs better than 4\,KB pages with an average 0.94 and 0.96 precision, and 0.92 and 0.90 recall for the moderate and aggressive variants respectively. \ptp achieved an average precision of 0.96 and recall of 0.97 with both variants.

As discussed in \S\ref{motivation:dis}, we expect the efficiency of \damon with huge pages to drop as the memory footprint increases to several terabytes. Hence using huge pages does not fundamentally solve the telemetry inefficiencies with \damon. Nevertheless, as we show in \S\ref{subsubsec:overheads}, \ptp outperforms \damon in terms of computational overheads even when huge pages are used.

\subsubsection{Sub-terabyte (SubTB) workloads}
The goal is to (i) test the capability of \ptp to identify hot regions even for low memory footprint or gigabyte-scale workloads and (ii) to demonstrate the memory footprint threshold at which DAMON and PMU starts deteriorating. 
We configure MASIM to allocate 1\,GB, 10\,GB and 100\,GB of heap and perform random loads within a 10\% hot region.

\begin{figure}[htb]
    \centering

    \includegraphics[width=0.8\linewidth]{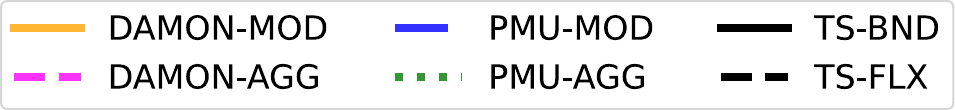}
    
    \subfloat[SubTB-1GB]{%
        \includegraphics[width=0.49\linewidth]{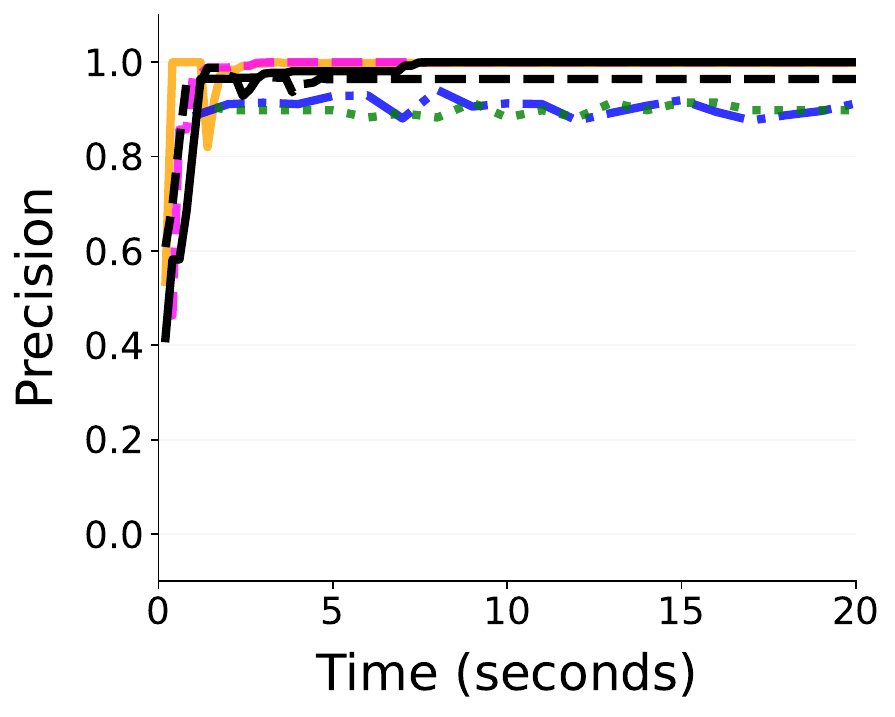}
        \includegraphics[width=0.49\linewidth]{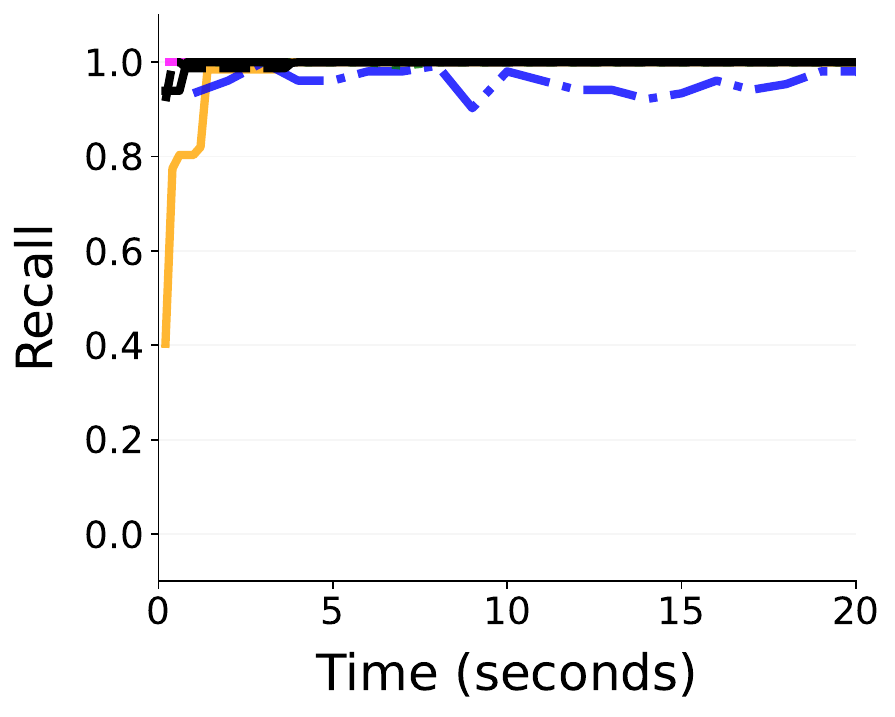}
    }

    \subfloat[SubTB-10GB]{%
        \includegraphics[width=0.49\linewidth]{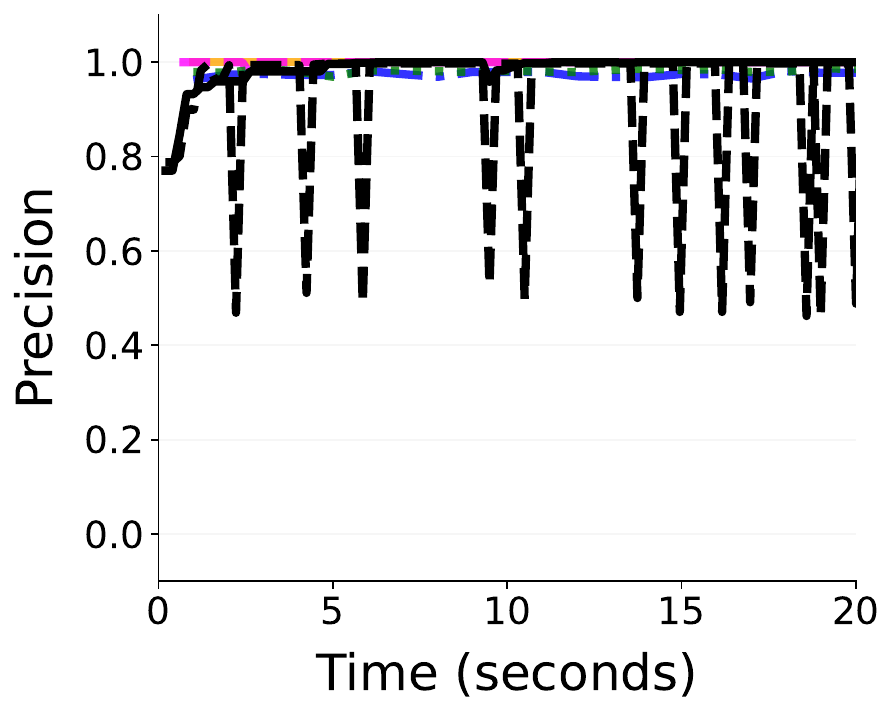}
        \includegraphics[width=0.49\linewidth]{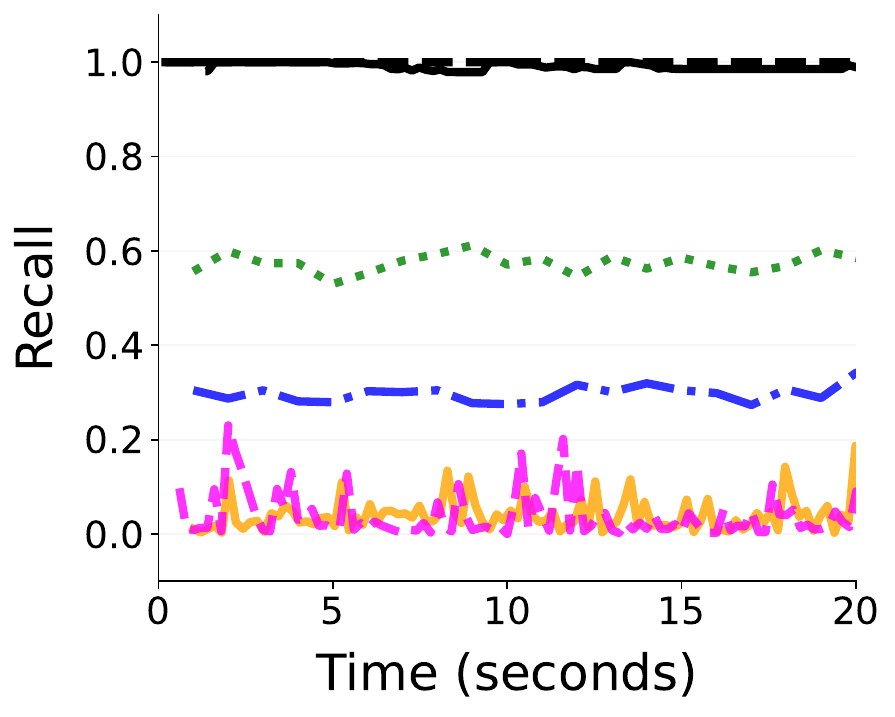}
    }

    \subfloat[SubTB-100GB]{%
        \includegraphics[width=0.49\linewidth]{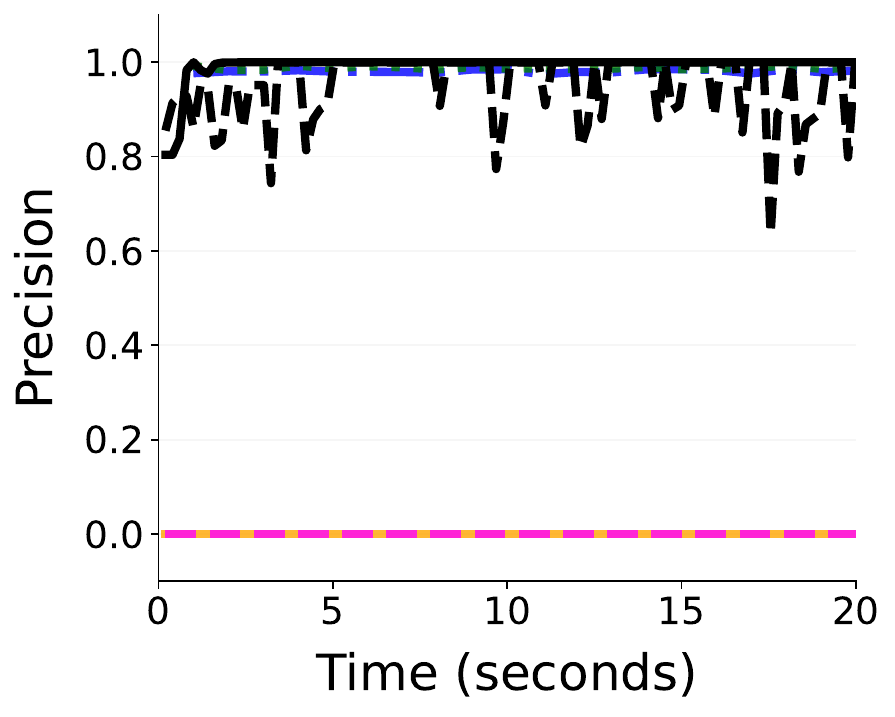}
        \includegraphics[width=0.49\linewidth]{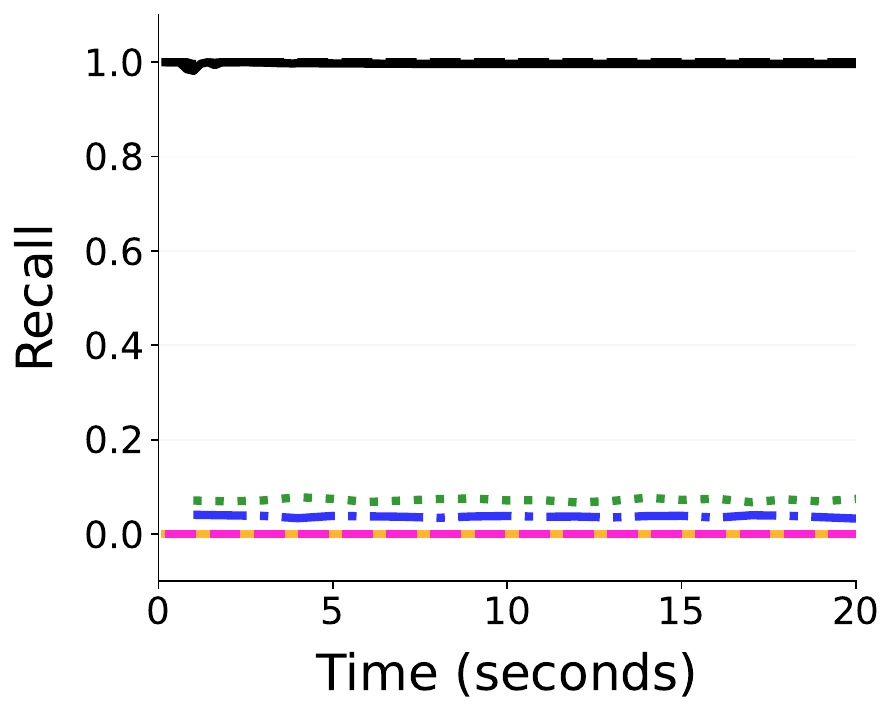}
    }
    \caption{Precision and recall for \textit{sub-terabyte} microbenchmark}
    \label{fig:accuracy-coverage}
\end{figure}

\smallskip
\noindent \textbf{Precision and recall.}
Figure~\ref{fig:accuracy-coverage} shows precision and recall plots for the SubTB workloads. 
For 1\,GB workload, \damon and PMU achieve a steady state of 0.9 and above for both precision and recall within a few seconds into the benchmark execution. But at 10\,GB, the recall drops significantly for both the variants of \damon. For 100\,GB workload, \damon's precision and recall drop to zero. 
PMU's precision remains high as they include data only from actual event samples.
However, as the size of the hot region increases the hot data coverage drops significantly (as discussed before in limitations of hardware counters, \S\ref{sec:hardware_counters}). This clearly shows that both DAMON and PMU fail to precisely capture the hot data regions as we scale to large memory footprint applications.
Both variants of \ptp outperform both DAMON and PMU in all the scenarios. The precision for \ptp-FLX
momentarily drops in between but quickly recovers. This is because \ptp-FLX variant is flexible to go beyond the region’s bounds, but with a certain error threshold, while picking a page table level for profiling.

\subsubsection{Needle in a haystack}
The goal is to test the capability to identify hard-to-find small hot data regions in a large heap by having a small hot data region of 50\,MB in a 5\,TB heap. 
Both variants of DAMON completely fail to capture hot regions with a zero precision and recall. However, both variants of PMU capture the hot regions with 0.81 precision and 0.99 recall on an average.
\ptp also successfully captures the small hard-to-find hot regions with 0.88 and 0.92 precision and 0.88 and 0.92 recall on an average for \ptp-BND and \ptp-FLX, respectively.

\subsubsection{Performance overhead analysis}
\label{subsubsec:overheads}

We present the computational overheads incurred by the telemetry techniques for the microbenchmarks described above. 

\begin{figure}
    \centering
    \includegraphics[width=.9\linewidth]{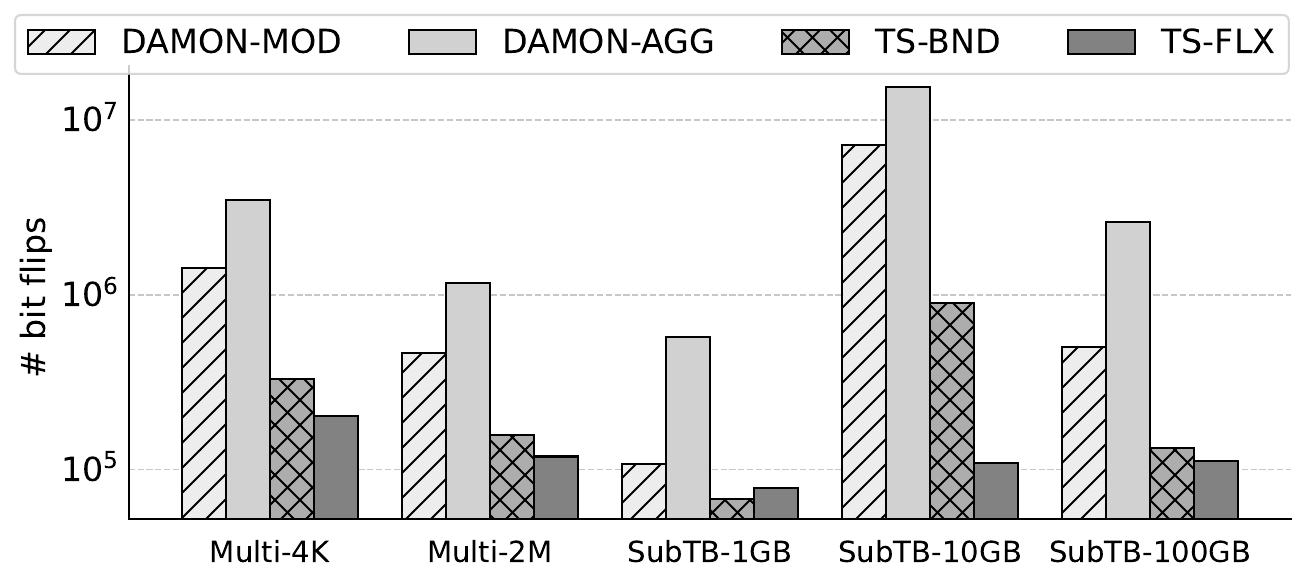}
    \caption{Total number of \accessed bit flipped by \ptp and DAMON. Y-axis is in log scale}
    \label{fig:bitflips}
\end{figure}

\begin{table}
  \footnotesize
    \centering
    \caption{Cycles (in billions) consumed by the kernel thread for DAMON and \ptp.}
    \label{tab:kdamond-cycles}
    \begin{tabular}{|c|C{.72cm}|C{.72cm}|C{.9cm}|C{.9cm}|C{.9cm}|}
        \hline
        \textbf{Config.}& \textbf{Multi-4K} & \textbf{Multi-2M} & \textbf{SubTB-1GB} & \textbf{SubTB-10GB} & \textbf{SubTB-100GB}  \\
        \hline
       {DAMON-MOD} & 9.55 & 2.42  & 1.15 & 19.53 & 3.52 \\ \hline
{DAMON-AGG} & 24.27 & 11.94 & 5.80 & 68.22 & 18.91 \\ \hline
{\ptp-BND} & 2.25 & 2.09  & 0.83 & 3.20 & 1.27 \\ \hline
{\ptp-FLX} & 2.28 & 1.80  & 0.95 & 1.16 & 1.19 \\ \hline
    \end{tabular}
\end{table}


\sepblock
\noindent \textbf{Bit flips.}
The number of \accessed bits flipped by \ptp is significantly less than \damon in all the microbenchmarks as shown in Figure~\ref{fig:bitflips}. This is because, in most cases, a single access bit at the higher levels of the page table tree is sufficient to cover a significant portion of a region.

\sepblock
\noindent \textbf{Computational overheads.}
Table~\ref{tab:kdamond-cycles} shows the number of cycles consumed by the kernel thread that performs the profiling for both variants of DAMON and \ptp. 
It can be observed that for all the microbenchmarks, both variants of \ptp consume significantly fewer CPU cycles.

In addition, the average single CPU utilization of the kernel thread for both variants of \ptp is 0.009\%, while it is 0.033\% and 0.09\% for \damon-MOD and \damon-AGG (significantly less than the 2.78\%--49\% CPU utilization incurred by linear scanning (Figure~\ref{fig:time_taken})). PMUs have been excluded from this comparison as the profiling is taken care of by the hardware and not in a separate kernel thread.

\begin{figure}
    \centering
    \includegraphics[width=.9\linewidth]{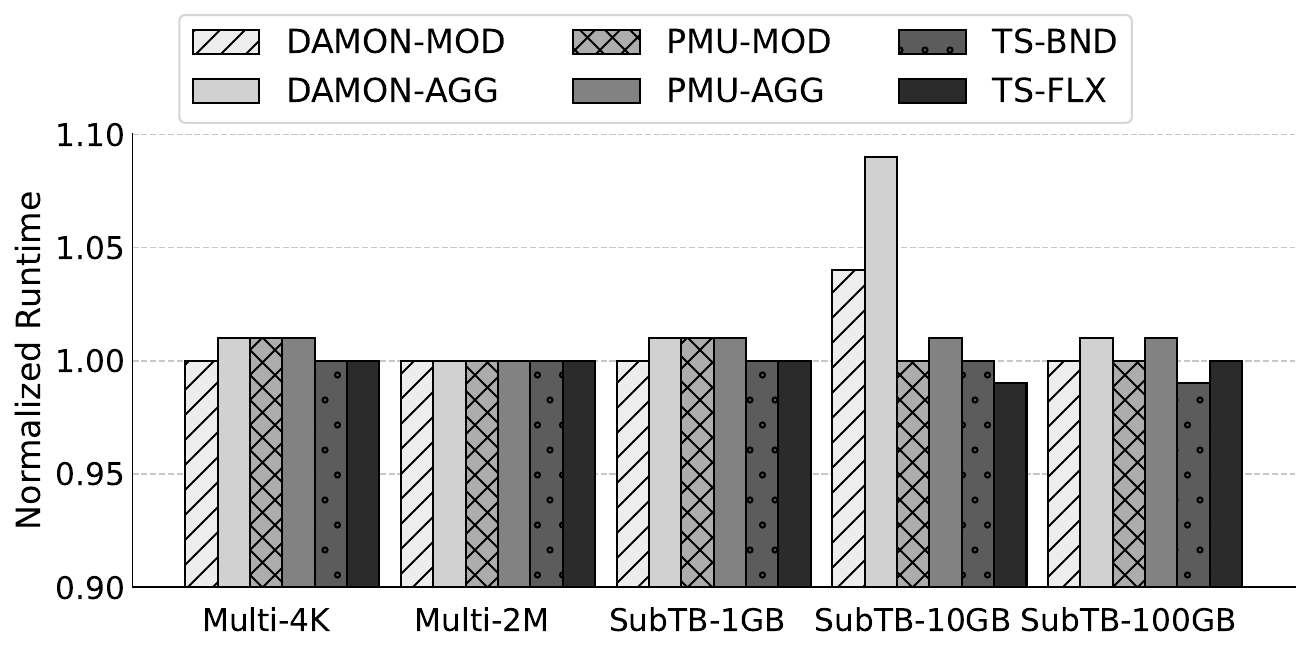}
    \caption{Impact on benchmark runtime with DAMON, \ptp, and PMU normalized to baseline with telemetry disabled. }
    \label{fig:slowdown}
\end{figure}

\sepblock
\noindent \textbf{Runtime impact.}
Figure~\ref{fig:slowdown} shows the execution time impact on the benchmarks, which excludes the memory initialization phase. Values are normalized to the baseline run where telemetry is disabled. We do not migrate any pages to measure pure telemetry overheads. 
It can be observed that \ptp does not impact the runtime of the microbenchmarks. PMU-AGG and DAMON-AGG impact the runtime of the microbenchmark in a few cases.

\subsection{Real-world application benchmarks}
\label{subsec:real-world}

\begin{figure*}
\centering
    \subfloat[\memcached with \ycsb\label{fig:memcached_ycsb_thp}]{%
        \includegraphics[width=0.48\textwidth]{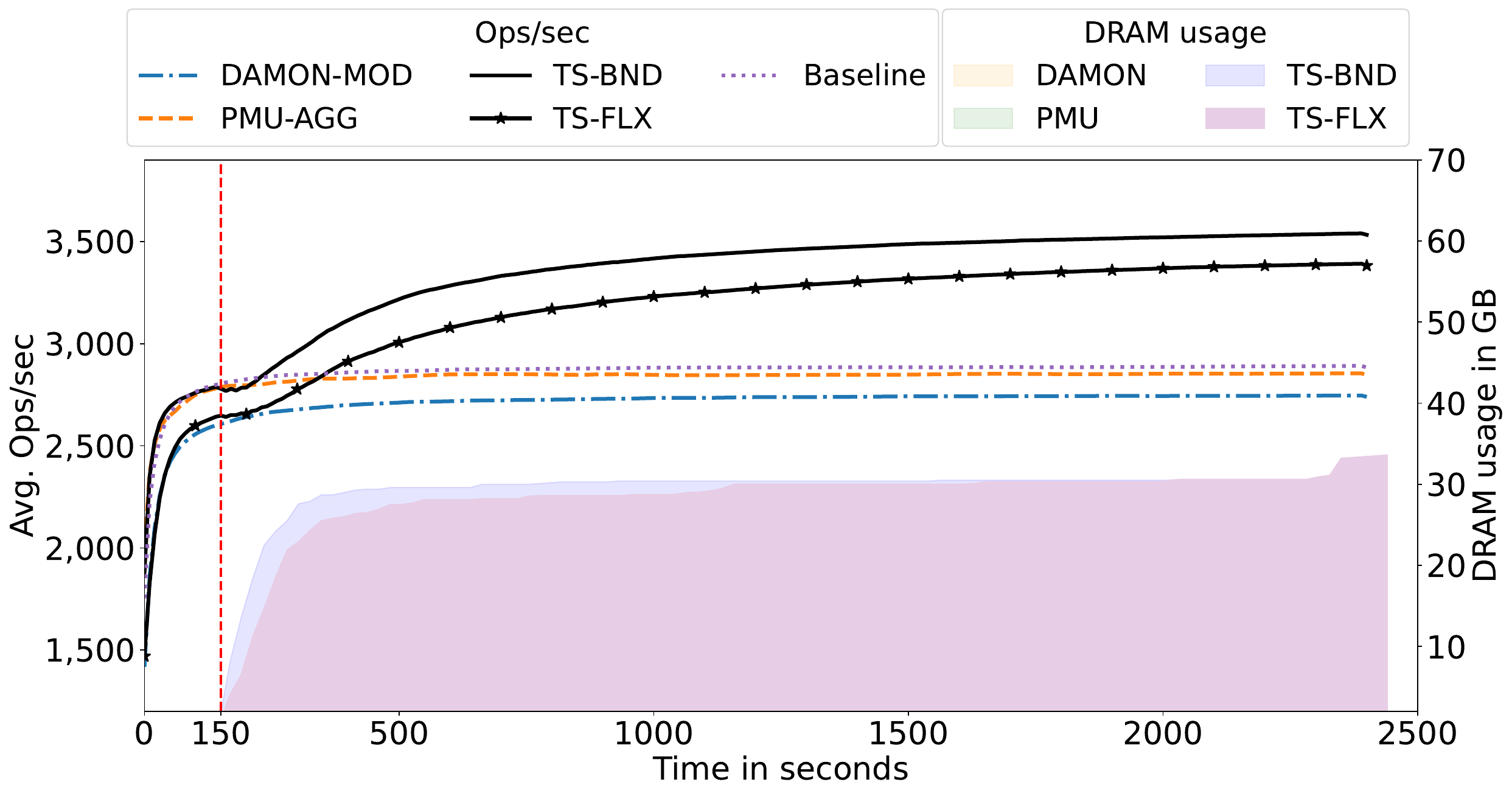}
    }
    \hfill
    \subfloat[\redis with \ycsb\label{fig:redis_ycsb_thp}]{%
        \includegraphics[width=0.48\textwidth]{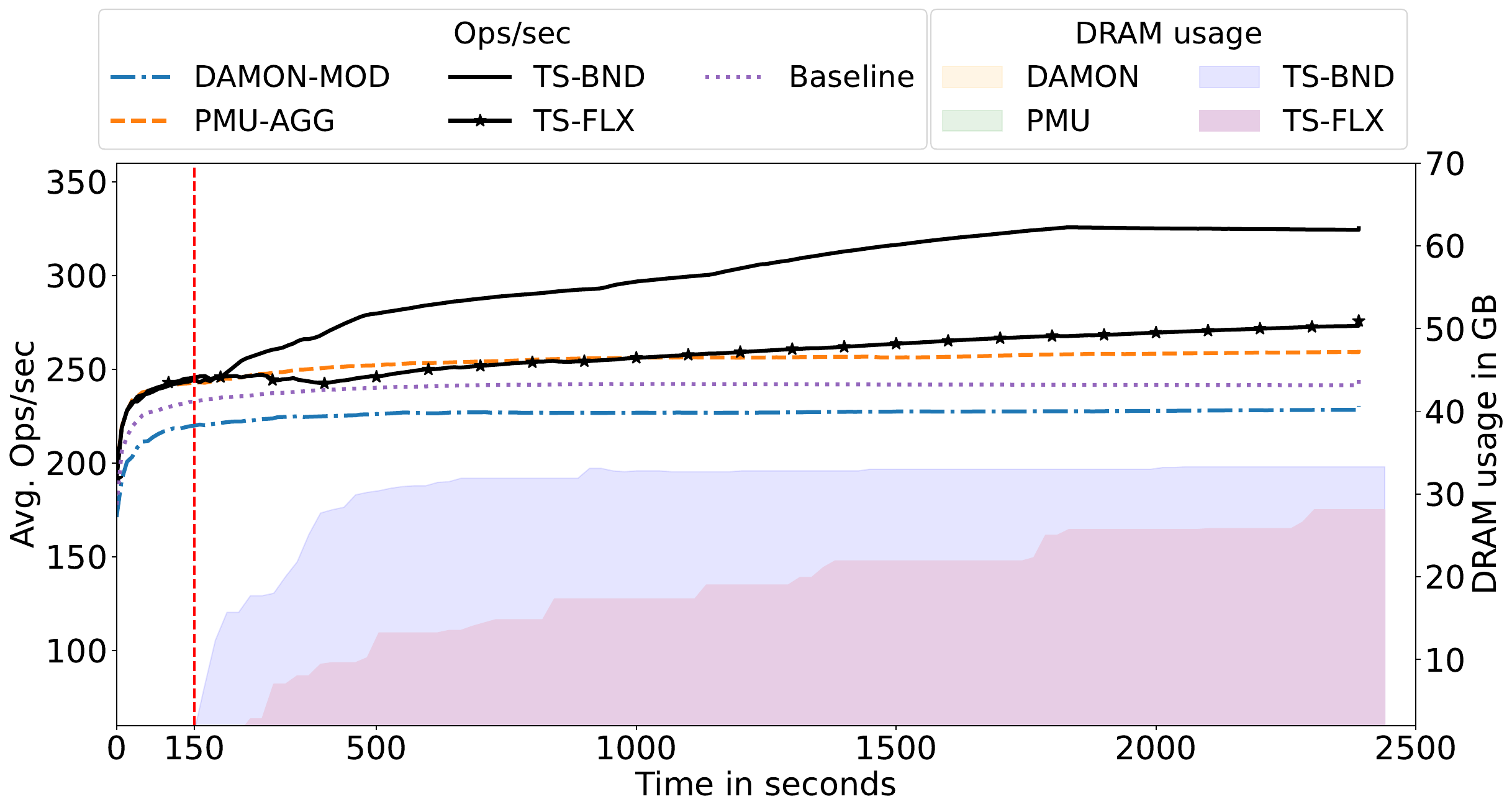}
    }

    \subfloat[\memcached with \memtier\label{fig:memcached_memtier_thp}]{%
        \includegraphics[width=0.48\textwidth]{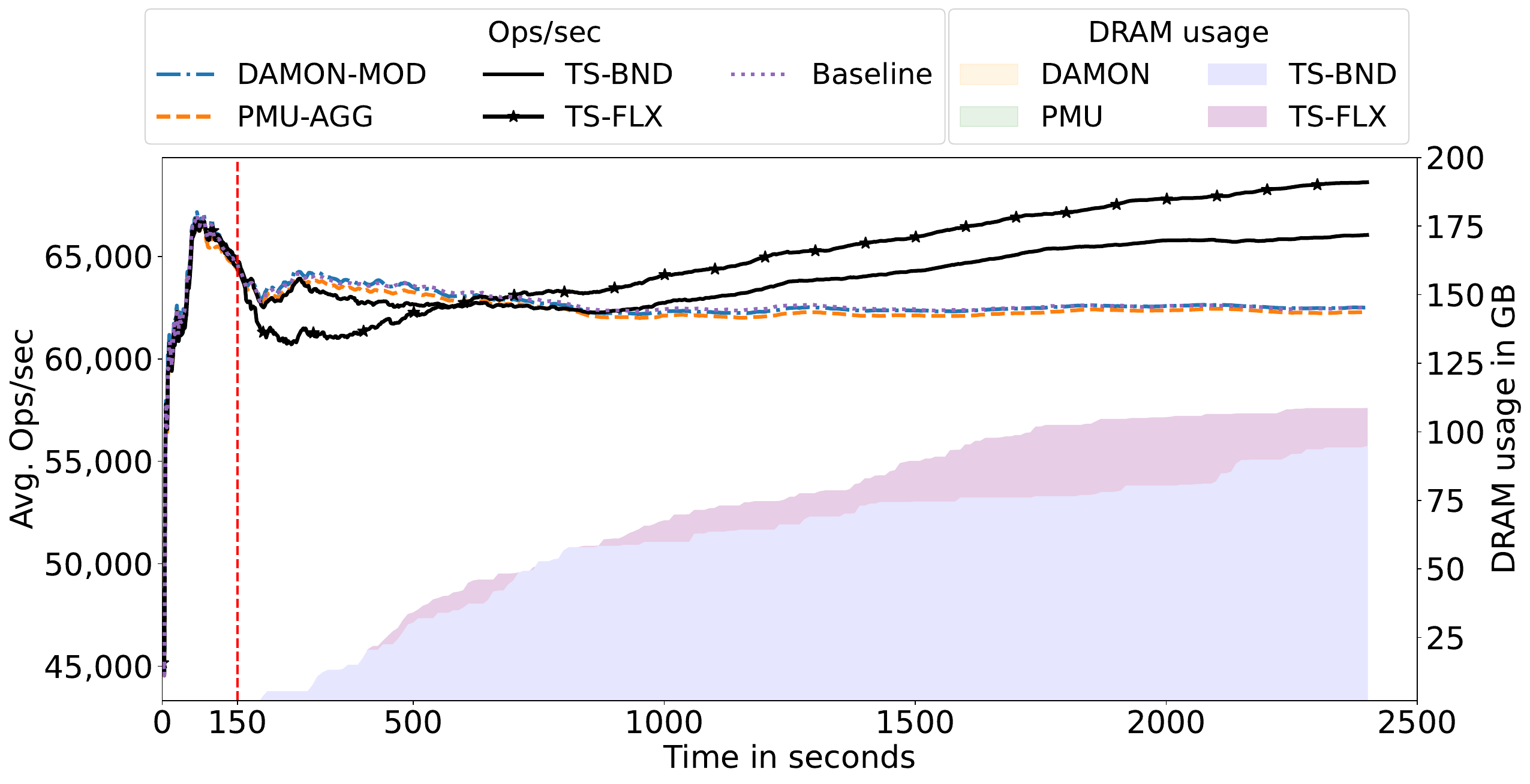}
    }
    \hfill
    \subfloat[\redis with \memtier\label{fig:redis_memtier_thp}]{%
        \includegraphics[width=0.48\textwidth]{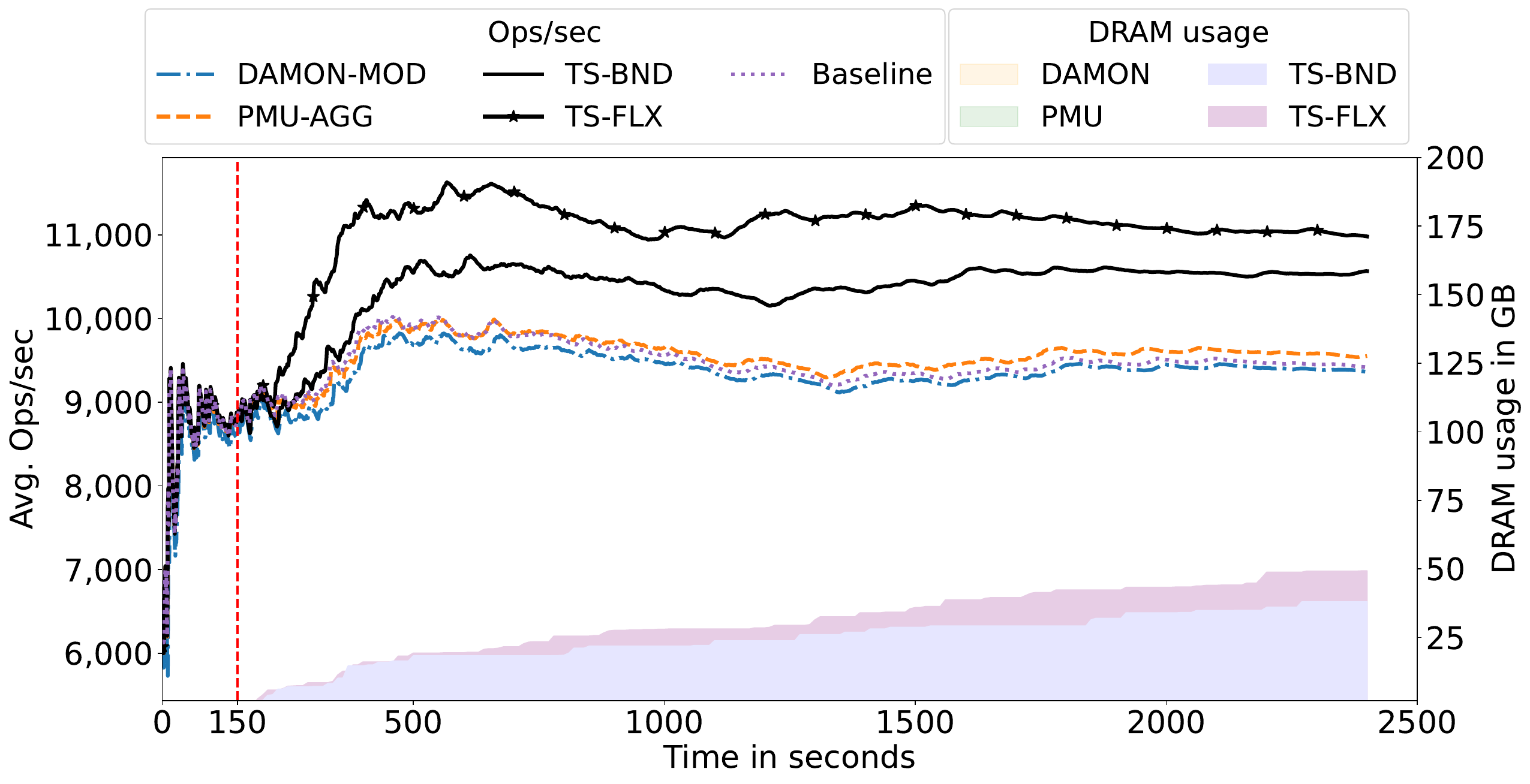}
    }
    \caption{Throughput improvement and DRAM usage for \memcached and \redis as hot data pages are migrated from far memory tier to near memory tier after a warmup period of 150 seconds.}
    \label{fig:throughput}
\end{figure*}

\begin{table}
  \centering
  \footnotesize
  \caption{\memtier~\cite{memtier-github} and \ycsb~\cite{ycsb} configurations.}
  \label{tab:memtier_config}
  \begin{tabular}{|r|L{2.5cm}|L{2.5cm}|}
    \hline
    \textbf{Parameter}   & \textbf{Memtier (MT)~\cite{memtier-github}} & \textbf{YCSB~\cite{ycsb}} \\ \hline
    Memory footprint & 1\,TB & 2\,TB \\ \hline
    Number of keys       & 200K                                   & 1000\,K                    \\ \hline
    Key Value size       & 5\,MB                                  & 2\,MB                     \\ \hline
    Number of threads    & 170                                     & 170                        \\ \hline
    Execution time                 & 40\,mins                                 & 40\,mins                        \\ \hline
     Hot data distribution & Gaussian w/ Std. deviation 100                              & Hotspot (99\% ops on 1\% hot data)                  \\ \hline
  \end{tabular}
\end{table}

In this section, we present the results for large memory footprint real-world applications on a tiered memory system.
We use the widely used \textit{Memcached}~\cite{memcached}, a commercial in-memory object caching system, and \textit{Redis}~\cite{redis}, a commercial in-memory key-value store, as our real-world application benchmarks. We use Memtier~\cite{memtier-article,memtier-github} and YCSB~\cite{ycsb} that generate different access patterns 
as our load generators~\cite{memtier-usage-1,memtier-usage-2,ycsb} for both Memcached and Redis to have a total of four real world scenarios. We configure the load generators as shown in Table~\ref{tab:memtier_config}. 

\subsubsection{Experiment setup}
We initialize the data on the far memory tier (Optane NVM) using interleaved memory allocation policy~\cite{numactl}. 
Once the data is initialized, we execute the workloads for 40\,minutes each. The first 150 seconds is the warmup phase, after which we start the telemetry technique to identify and migrate hot data from the far memory tier to the near memory tier. 
We compare the performance improvement over baseline (telemetry disabled) with \damon, \hemem, and \ptp. 

\subsubsection{Hot page classification and migration.}
\label{sec:hot_page_detection}
The information generated by a typical telemetry technique is a set of pages (or regions), access timestamp, and the number of times they were accessed in a time window.
Whether a particular page or region is hot or not is up to the user to define based on the application's behavior and requirements~\cite{hemem, hmkeeper,tpp}. In addition, migrating pages across memory tiers have associated overheads and hence should be rate limited. 

We use the following rules to classify and migrate hot pages (or regions) as also used in prior works~\cite{hemem, hmkeeper,tpp}:  
\one we consider regions with access count greater than a threshold (set to 5) as hot,
\two we skip large regions ($\ge$ 4\,GB) to ensure hot pages are migrated at a finer granularity. Subsequent profiling windows split larger hot regions, and hence they are eventually migrated,
\three for the rest of the regions, we start migrating regions with the highest hotness score and stop once a limit of 10\,GB is reached.

\newcommand{\rowsize}{0.76cm}
\newcommand{\tworowsize}{2.6cm}

\begin{table}
  \centering
  \caption{Latency impact and data pages migrated with different telemetry techniques.}
  \label{tab:compare_and_improvement}
  {%
    \footnotesize
      \begin{tabular}{|l|l|c|c|c|c|}
      \hline
    
  & \textbf{Config.} & \multicolumn{2}{c|}{\textbf{95th \%tile lat. (ms)}} & \multicolumn{2}{c|}{\textbf{Data migrated (GB)}} 
      \\   \hline
      \multirow[c]{5}{*}[0em]{\rotatebox{90}{\textbf{Memcached}}}& &YCSB & MT & YCSB & MT\\ \cline{2-6}
            &\textbf{DAMON-MOD }          & 881    &  11.2    & 0     &   0    \\\cline{2-6}
      &\textbf{PMU-AGG}    & 976       & 11.3 & $\approx$0.01    &  $\approx$0       \\\cline{2-6}
      &\textbf{\ptp-BND}    & 867      & 10.8 & $\approx$31    & $\approx$94        \\\cline{2-6}
      &\textbf{\ptp-FLX}      & 824      &  10.5  & $\approx$34    & $\approx$108   \\
      \hline \hline
      \multirow[c]{4}{*}{\rotatebox{90}{\textbf{\redis}}}
      &\textbf{DAMON-MOD}        &850  & 59.13& 0 &  0   \\ \cline{2-6}
      &\textbf{PMU-AGG}    & 757    &    57.5   & $\approx$0.15   &    $\approx$0.05   \\ \cline{2-6}
      &\textbf{\ptp-BND}      & 696   &   54.01    & $\approx$34   &   $\approx$38    \\ \cline{2-6}
      &\textbf{\ptp-FLX}        & 741      &  55.55  & $\approx$28   &  $\approx$50     \\ \hline
      
    \end{tabular}
  }
  
\end{table}

\subsubsection{Results}
\label{sec:results}
\begin{figure}
    \centering
    \includegraphics[width=.9\linewidth]{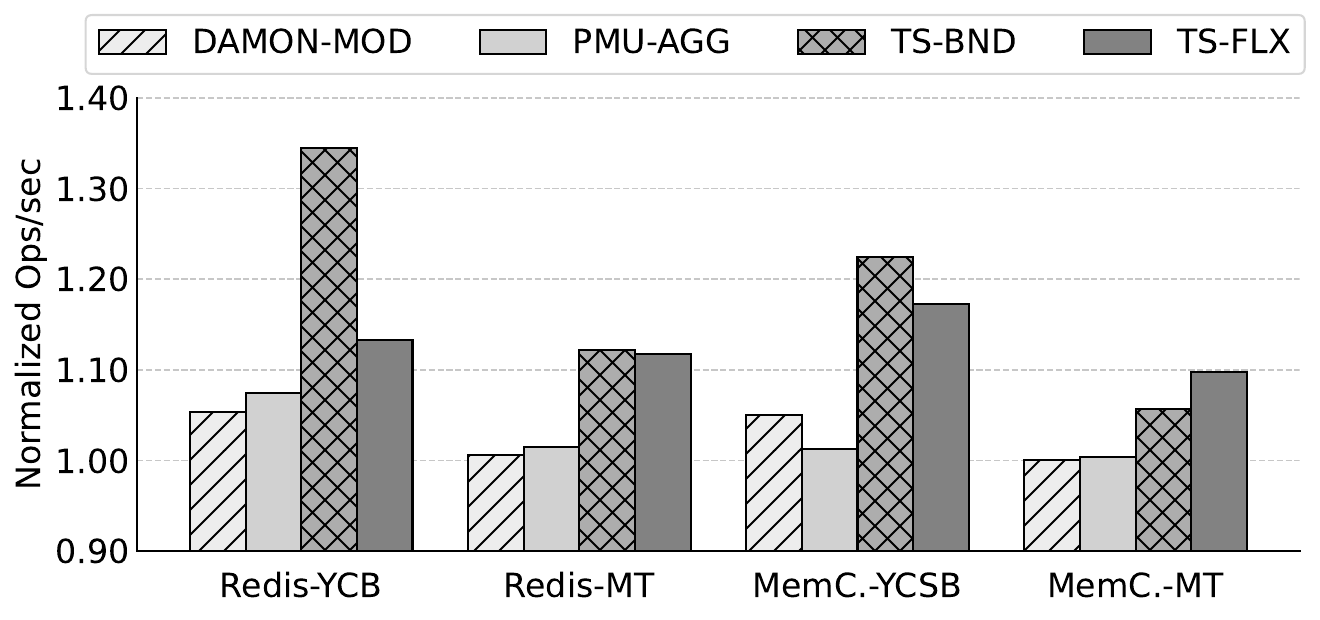}
    \caption{Normalized throughput improvement for \memcached \redis.}
    \label{fig:clear_perf_imp}
\end{figure}

As shown in Figure~\ref{fig:throughput}, 
\damon could not identify a single hot data page, but the profiling overheads resulted in decreased throughput compared to baseline. \hemem identified only 157\,MB of hot data (Table~\ref{tab:compare_and_improvement})  out of a few gigabytes of hot data set and hence resulted in marginal throughput improvement in some cases. Both variants of \ptp detected and migrated significant portion of the hot data pages to near memory tier resulting in up to 34.4\% throughput improvement compared to baseline with both telemetry and page migration disabled (Figure~\ref{fig:clear_perf_imp}). In addition, Table~\ref{tab:compare_and_improvement} shows 
latency values where \ptp outperforms both \damon and \hemem. 

%

\section{Conclusion}
\label{sec:conclusion}

Tiered memory architectures offer an attractive way to provide high memory capacity efficiently.
Precise and timely telemetry is critical for proactive hot and cold data placement in the appropriate tiers.
\ptp is a novel technique based on page table profiling that meets the telemetry requirements of a tiered memory system that can scale for terabyte-scale applications and that is also portable across different hardware architectures.

\bibliographystyle{plain}
\bibliography{refs}

\end{document}